\documentclass[prd,showpacs, twocolumn, nofootinbib, preprintnumbers]{revtex4}%

\usepackage{amssymb,hyperref, amsmath}
\usepackage[dvips]{color}
\usepackage[dvips]{graphicx}
\usepackage{epsfig}

\def \emu {\epsilon^\mu}
\newcommand{\pfdotx}{\vec{p}_f \cdot \vec{x}}
\newcommand{\pfdotz}{\vec{p}_f \cdot \vec{z}}

\begin{document}

\preprint{JLAB-THY-06-457}

\title{Radiative Transitions in Charmonium from Lattice QCD}

\author{Jozef J. Dudek}
\affiliation{Jefferson Laboratory MS 12H2, 12000 Jefferson Avenue, Newport News, VA 23606, USA}
\email{dudek@jlab.org}

\author{Robert G. Edwards}
\affiliation{Jefferson Laboratory MS 12H2, 12000 Jefferson Avenue, Newport News, VA 23606, USA}
\email{edwards@jlab.org}

\author{David G. Richards}
\affiliation{Jefferson Laboratory MS 12H2, 12000 Jefferson Avenue, Newport News, VA 23606, USA}
\email{dgr@jlab.org}

\begin{abstract}
Radiative transitions between charmonium states offer an insight into
the internal structure of heavy-quark bound states within QCD. We
compute, for the first time within lattice QCD, the transition
form-factors of various multipolarities between the lightest few
charmonium states. In addition, we compute the experimentally
unobservable, but physically interesting vector form-factors of the
$\eta_c, J/\psi$ and $\chi_{c0}$.

To this end we apply an ambitious combination of lattice techniques,
computing three-point functions with heavy domain wall fermions on an
anisotropic lattice within the quenched approximation. With an
anisotropy $\xi=3$ at $a_s \sim 0.1 \, \mathrm{fm}$ we find a
reasonable gross spectrum and a hyperfine splitting $\sim 90
\mathrm{MeV}$, which compares favourably with other improved actions.

In general, after extrapolation of lattice data at non-zero $Q^2$ to
the photopoint, our results agree within errors with all well measured
experimental values. Furthermore, results are compared with the
expectations of simple quark models where we find that many features
are in agreement; beyond this we propose the possibility of
constraining such models using our extracted values of physically
unobservable quantities such as the $J/\psi$ quadrupole moment.
 
We conclude that our methods are successful and propose to apply them
to the problem of radiative transitions involving hybrid mesons, with
the eventual goal of predicting hybrid meson photoproduction rates at
the GlueX experiment.

\end{abstract}


\maketitle %%% typeset front matter (including abstract)

\section{Introduction}\label{intro}

Charmonium occupies a valuable intermediate position within QCD, being
neither in the purely non-relativistic regime nor the regime where
chiral symmetry breaking dominates the physics. This makes it a
relatively clean system in which to study non-perturbative QCD
dynamics, and indeed QCD-inspired quark-potential models as well as
lattice QCD have been rather successful in describing the observed
features of the spectrum\cite{Brambilla:2004wf}. However, charmonium
cannot be considered to be completely understood; as an example, in
recent years a number of new charmonium resonances have been claimed
in experiment (see \cite{Swanson:2006st} for a review), several of
which cannot be easily reconciled with the predictions of simple
quark-potential models.

Unlike the light quark sector, in charmonium the lightest state for
most $J^{PC}$'s lies below the threshold for OZI-allowed decay and
consequently these states are rather narrow. These states have been
the subject of many calculations in lattice QCD which generally
reproduce the gross structure of the spectrum, but are unable to
account for all of the detailed fine structure (such as the 117 MeV
$J/\psi - \eta_c$ splitting), owing to some combination of the
approximations inherent in the computations, which can include the
finite lattice spacing, quenching and lack of disconnected diagrams.

Masses are not the only well measured charmonium observables. Because
of the small total width of these lightest few states, radiative
transitions between them constitute large branching fractions and have
been measured experimentally by a number of groups
\cite{PDBook,Adam:2005uh}. These quantities have been studied within
quark-potential models (and latterly EFT approaches like pNRQCD) where
they are related to the overlap of meson wavefunctions with the photon current operator and as such are an insight into the
internal structure of these states. QCD sum rules have also been
applied with some success\cite{Beilin:1984pf} . No study of radiative transitions
in charmonium has yet been performed using lattice QCD - it is this
situation that we remedy in this paper.

This study in the charmonium sector is an ideal test-bed for our
eventual aim of computing photocouplings in the light-quark sector, in
particular the coupling between a conventional meson, a photon and a
hybrid meson. Such a coupling drives the photoproduction mechanism
proposed by the GlueX collaboration for their experiment in Hall~D of
the upgraded CEBAF at Jefferson Lab\cite{Dzierba:2001wg}. Flux-tube
model calculations \cite{Close:2003ae, Close:2003fz} suggest that
these couplings are not small and that we may expect copious
production of hybrid mesons, but the assumptions underlying such a
model need to be tested in a framework closer to QCD. This paper will
focus on transitions between conventional charmonium mesons, and we
will proceed to transitions to hybrid mesons in a subsequent work.

In section \ref{sim} we outline the computational details of our
lattice calculation and in section \ref{spectrum} we present the
charmonium spectrum so obtained. In section \ref{three} we explain how
three-point functions extracted from the computation can be converted
into multipole form-factors and in section \ref{ff} we present lattice
estimates for the vector form-factors of the lightest three charmonium
states and compare with what one would expect on the basis of simple
quark models. In section \ref{trans} we consider radiative transitions
and discuss the extrapolation from the non-zero $Q^2$ accessible on
our finite-volume lattice to the relevant case of an on-shell
photon. Finally in section \ref{discuss} we conclude, relegating the
details of Lorentz covariant multipole decompositions and scale
setting on anisotropic lattices to appendices. 

\section{Computational Details}\label{sim}

The computations were performed in the quenched approximation to QCD,
using the Chroma software system~\cite{Edwards:2004sx}.
We employed 300 configurations on a $12^3 \times 48$ lattice generated
using an anisotropic Wilson gauge action~\cite{Klassen:1998ua}, with a
renormalized anisotropy $\xi \equiv a_s/a_t = 3$.  The temporal
lattice spacing obtained from the static quark-antiquark potential is
$a_t^{-1} = 6.05 (1)~{\rm GeV}$. 

Anisotropic lattices as applied to charmonium exploit the fact that
while the quark mass scale demands a cut-off above $\sim 1.5~
\mathrm{GeV}$, the internal three-momentum scale is typically much lower,
$\sim 500~\mathrm{MeV}$. On our lattice, we can have both $m_c a_t$
and $| \vec{p} | a_s$ reasonably small and a spatial
length $\gtrsim 1~\mathrm{fm}$ without requiring very many spatial
lattice sites. We work on only one volume, $L_s \approx 1.2~
\mathrm{fm}$; previous charmonium spectrum studies indicate that
there are no significant finite volume effects for lattices of this
size or larger\cite{Choe:2003wx,Drummond:1999db}. With this volume and anisotropy the three-momenta accessible are $\vec{p} = \tfrac{2 \pi}{\xi L_s a_t}(n_x, n_y, n_z) \approx 1.06\, \mathrm{GeV} (n_x, n_y, n_z) $.

The quark propagators were computed using an anisotropic version of the
domain-wall fermion (DWF) action~\cite{Shamir:1993zy} with a
domain-wall height $a_t M = 1.7$, a fifth dimensional extent $L_5 = 16$,
and a quark mass $a_t m_q$.
The conventions used for the action are defined in Ref.~\cite{Edwards:2000qv}.
In terms of dimensionless variables $\breve \psi = a_s^{3/2} \psi$ and
$\breve W_\mu = a_\mu W_\mu$, where $a_0 = a_t$ and $a_k = a_s, k = 1,2,3$.
The kernel of the domain action is the anisotropic Wilson fermion 
action~\cite{Klassen:1998fh,Chen:2000ej}:
\begin{equation}
a_t Q = - a_t M + \nu_t \breve W_t \gamma_4 + \frac{\nu_s}{\xi_0} \sum_k
 \breve W_k \gamma_k,
\label{eq:Q_Tim}
\end{equation}
where the Wilson operator
\begin{equation}
W_\mu \equiv \nabla_\mu - \frac{a_\mu}{2} \gamma_\mu \Delta_\mu ~.\nonumber
\end{equation}
The parameter $\xi_0$ in Eq.~\ref{eq:Q_Tim} is the bare anisotropy
which is determined in the gluonic sector so as to yield the desired
renormalized anisotropy $\xi$.
The remaining parameters of the action, $m_q,
\nu_s~\mbox{and}~\nu_t$ represent the quark mass, and the
renormalization of the couplings in the spatial and temporal
directions, respectively. The parameters $\nu_s$ and $\nu_t$ are not
independent; we will fix $\nu_t = 1$ and then tune $(m_q, \nu_s)$ so as
to yield the desired quark mass and such that the speed of light
obtained from the meson dispersion relations be $1.0$, as discussed in
the next section. 

While this computation is performed at only a single value of the
lattice spacing and hence no attempt can be made to determine the
lattice spacing dependence of the results, an important benefit to the
use of the domain-wall fermion action is that it is automatically
improved to ${\cal O}(a^2)$.

We anticipate that the quenched approximation will not be a serious
impediment to this study. In the sub-$D\bar{D}$-threshold charmonium system we
expect (on the basis of previous lattice studies) that the dominant
effect of light quarks will be to modify the running of the strong
coupling. In the quenched approximation the QCD $\beta$ function is
not properly reproduced, thus if we set our lattice scale using a long
distance dominated quantity (such as the $1P-1S$ splitting or some
intermediate distance in the static quark-antiquark potential) then we
will have a weakened coupling at short distances compared to
QCD. Hence within the quenched approximation we cannot expect to
correctly describe such short-distance-dominated quantities as the
hyperfine splitting, or meson decay
constants\cite{Gottlieb:2005me, Davies:2003ik}. Our principal interest is with
radiative transitions, which are dominated by long-distance
wavefunction overlaps, and hence should not be considerably distorted
by the quenched approximation.

\section{Spectrum}\label{spectrum}
Charmonium masses and interpolating field overlap factors were
extracted from fits to the {\em connected} two-point
functions\footnote{Disconnected diagrams will be discussed in section
\ref{three}}, $\Gamma^{(2)}_{ij}(\vec{p}; t) =$
\begin{equation} \label{twopt}
 \sum_{\vec{x}} e^{-i
   \vec{p}.\vec{x}} \langle \bar{\psi}\Gamma_i \psi(\vec{x}, t)  \;
 \bar{\psi}\Gamma_j \psi(\vec{0},0) \rangle =  \sum_N \frac{Z^{(N)}_i Z^{(N)*}_j}{2 E^{(N)}} e^{- E^{(N)}   t }
\end{equation}
The fact we used local fermion bilinears as interpolating fields 
limits us to the $J^{PC}$ listed in Table \ref{tab:two}.

In the previous section we discussed the need to have a rather fine
lattice spacing to ensure that charm quarks are not ``cut-off''; a
related problem arises from the two-scaledness of the charmonium
system. The scale which dictates our choice of lattice spacing is the
typical mass of a charmonium state $\sim 3 \mathrm{GeV}$, while the
level spacing e.g. $J/\psi - \psi'$ $\sim 600 \mathrm{MeV}$, is much
smaller. This second scale indicates how fast, relatively, the excited
state exponentials in eqn \eqref{twopt} fall off with $t$, which we
see will be rather slow, so that it is possible at finite times to get
contributions from many excited states. This is not the case with
light quarks where the level splitting is of the same size as the
ground state mass (e.g. $\rho(770), \rho(1450)$). Because many excited
states can be contributing during the finite temporal-extent of our
lattice, we are unlikely to reach a plateau and as such extracting the
ground-state mass is non-trivial.

We can reduce this problem somewhat if we improve the overlap of the
interpolating field on to the ground state by smearing it over
space. We do so using a gauge invariant cubic approximation to
a rotationally symmetric Gaussian,
\begin{multline}
\hspace{-5mm}  \left(1- \tfrac{3 \sigma^2}{2 N} \right)^N \left( 1 +
    \tfrac{\sigma^2/4N}{1- 3\sigma^2/2N} \sum_{i=1}^3 \left[ U_{x,i}
      \delta_{x, x+\hat{i}} + U^\dag_{x-\hat{i},i} \delta_{x,x-\hat{i}}
    \right]    \right)^N\\
  \xrightarrow[N \to \infty]{}  e^{-\sigma^2 \nabla^2/4}.\label{eq:gaussian}
\end{multline}
With the right smearing radius this will resemble the ground state
wavefunction and maximise the overlap, while the excited states, which
have radial nodes, will have a decreased overlap.

We computed two-point functions with smearing radii $\sigma = 0.0,
3.6, 4.0$. For the case $\sigma =
3.6$, we attempted to reduce gauge fluctuations by smearing the gauge
links entering into eqn.~\ref{eq:gaussian} using the gauge-invariant
procedure (APE) introduced in ref.~\cite{Albanese:1987ds}. 
For $\sigma=3.6, 4.0$ we used $N=32,50$ and $5,0$ APE smearing iterations. 

\begin{figure}[h]
  \centering
 
   \psfig{width=7.7cm,file=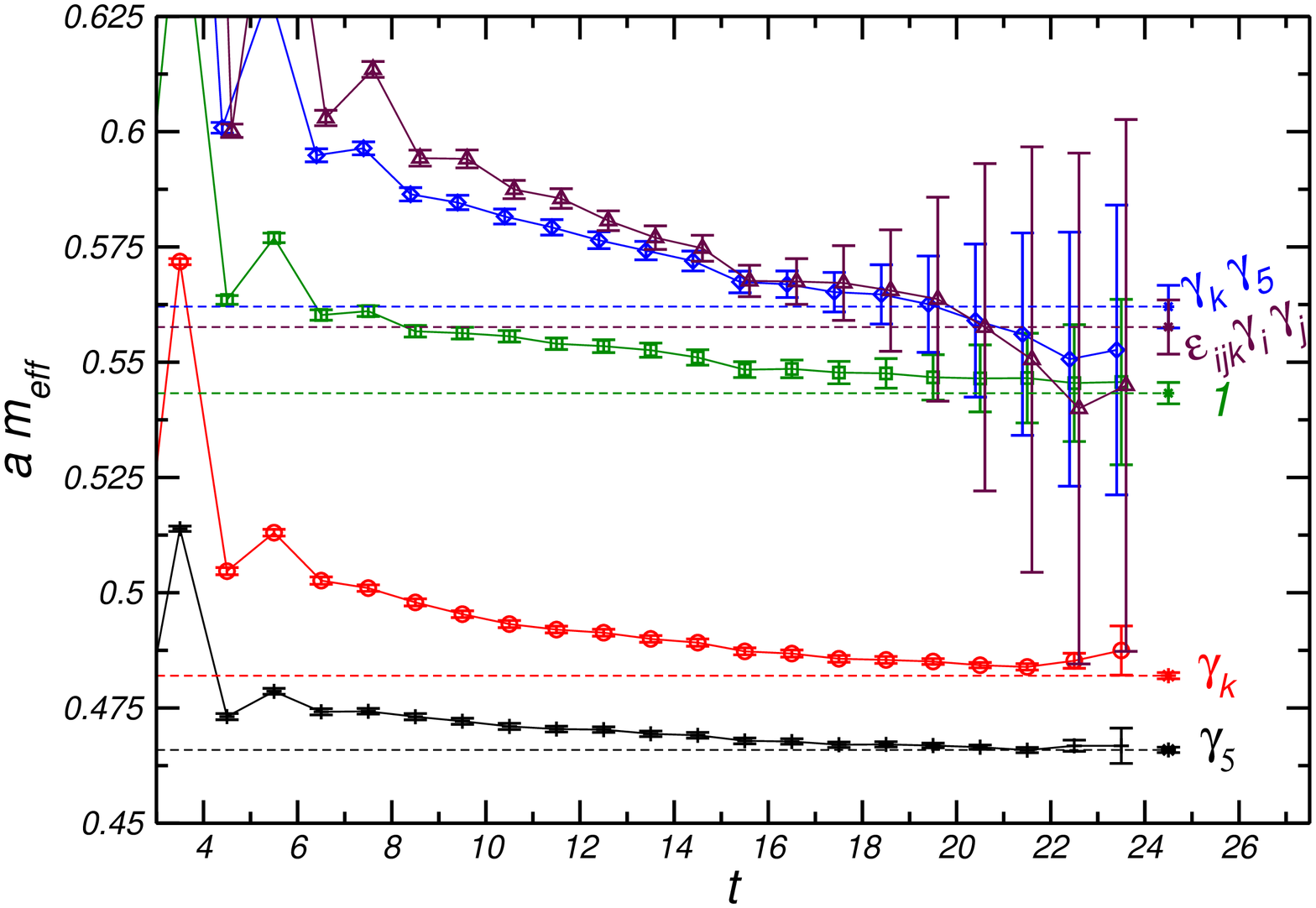}
      %\hspace{.5cm}

      \psfig{width=7.7cm,file= 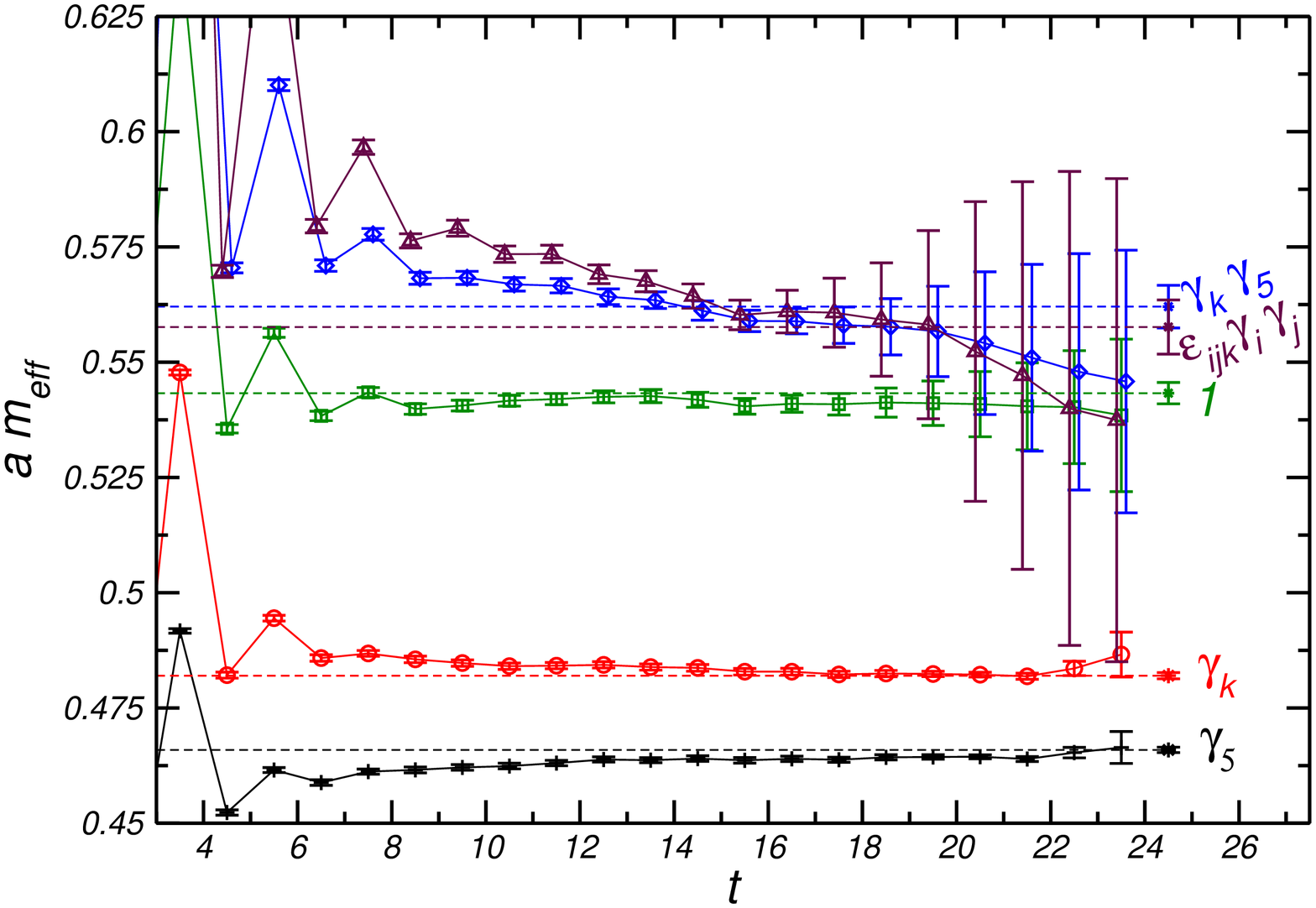}
       %\hspace{.5cm}

      \psfig{width=7.7cm,file= 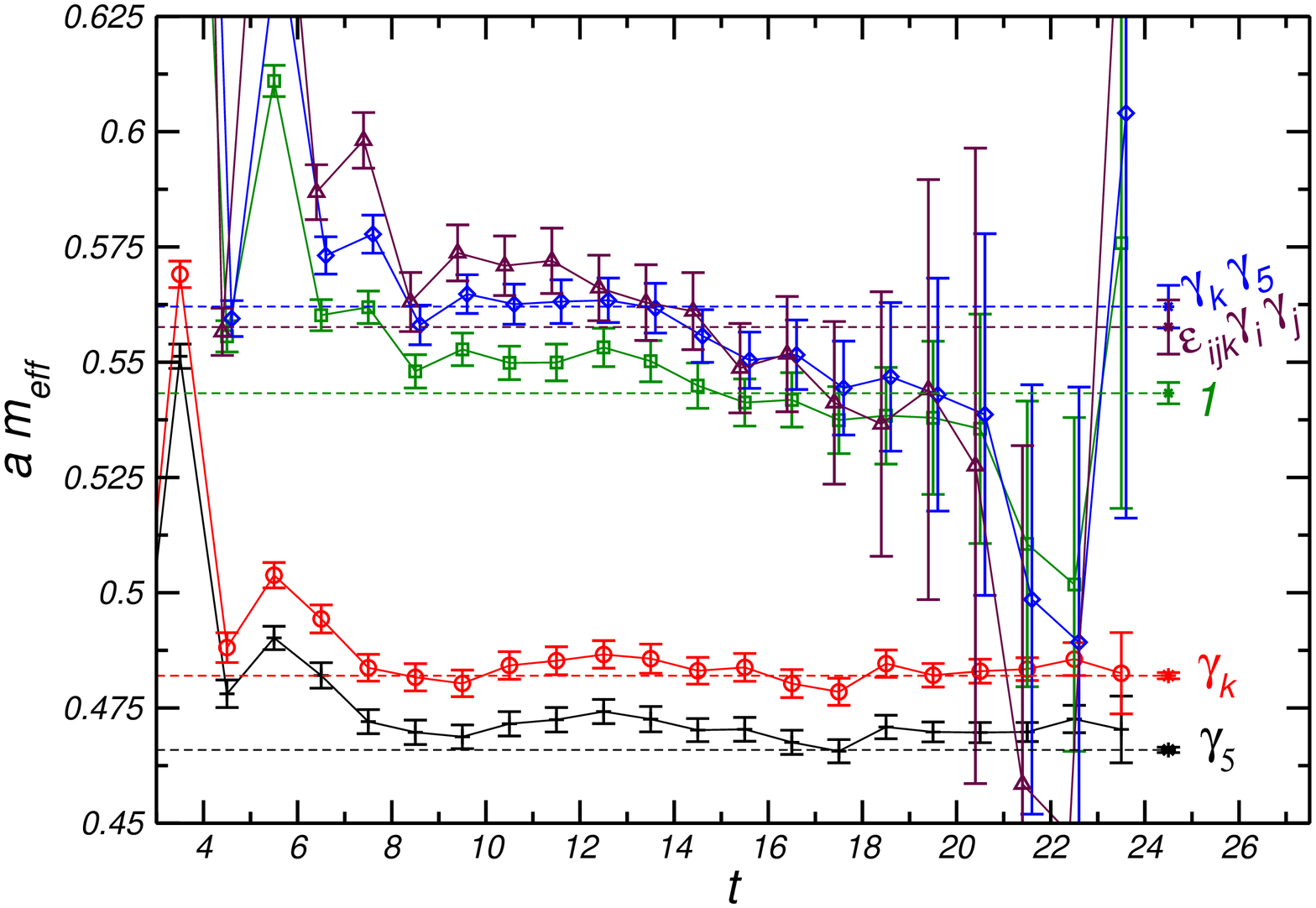}   
      \caption{Effective mass plots (at $\vec{p}=(000)$ using local
    interpolating fields. Dashed lines and rightmost data point show
    the ground-state mass obtained from the multi-correlator fit. (a)
    smeared($\sigma = 4.0$)-local (b) smeared($\sigma=3.6$)-local (c)
    smeared($\sigma=3.6$)-smeared($\sigma=3.6$) }
  \label{fig:meff}
\end{figure}

We display effective mass plots in figure \ref{fig:meff}, where the
oscillations in the first few timeslices are an artifact of the
non-locality of the DWF action in four dimensions.  The $\sigma=3.6$
smeared-smeared data has reasonable plateaux, at least for the
$\eta_c, \psi$ and $\chi_{c0}$.

Our fitting procedure is as follows.  Given a set of interpolating
operators $\{ O_i \}$, we can construct the matrix of time-sliced
  correlators
\begin{eqnarray}
\Gamma^{(2)}_{ij}(\vec{p}; t) & \equiv & \sum_{\vec{x}} e^{-i \vec{p} \cdot \vec{x}} \langle O_i(\vec{x}, t)
O_j^{\dagger}(0) \rangle
\nonumber\\
& = & \sum_N \frac{Z_i^{(N)}(\vec{p}) Z_j^{(N)*}(\vec{p})}{ 2 E^{(N)}} e^{-E^{(N)} t},\label{eq:corr_matrix}
\end{eqnarray}
where
\begin{equation}
Z_i^{(N)}(\vec{p}) = \langle 0 | O_i | N(\vec{p}) \rangle.\nonumber
\end{equation}
If we had computed every element of the $\Gamma^{(2)}_{ij}$, then we could
apply a variational method to diagonalise the system.  However, in general we have an
incomplete matrix, with several sink operators for each
source operator. Thus we employ a ``factorizing'' fit to this
incomplete matrix of correlators,
in which the fitted parameters are $\{ E^{(i)},
  Z^{(i)}_j: j = 1,\dots, N_{\rm op}, i = 1,\dots, N_{\rm exp} \}$,
  where there are $N_{\rm op}$ operators used in the fit, and we include
  $N_{\rm exp}$ states. We assume that the residues of the fit
  factorize according to eqn.~\ref{eq:corr_matrix} and that as
  demonstrated in Appendix \ref{Mink}, the $Z$'s are real. In practice we
  include the three smearing combinations shown in figure
  \ref{fig:meff} for each $J^{PC}$ and where possible fit to the
  ground state plus one excited state. For the $\chi_{c1}$ and $h_c$
  the noise on the data allowed for only inclusion of the ground state
  in the fit.

In order to propagate statistical error due to the finite number of
gauge field configurations through our calculations we adopt a single
elimination jackknife procedure. For the two-point function fits we
remove from the ensemble one configuration and average the remaining
two-point functions. The factorising fit is performed on this average
yielding a value for each of the fit parameters ($Z,E$). This is
repeated for each configuration yielding an ensemble for the fit
parameters over configurations. These are saved for later use in the
three-point function calculations. 

\begin{figure}[h]
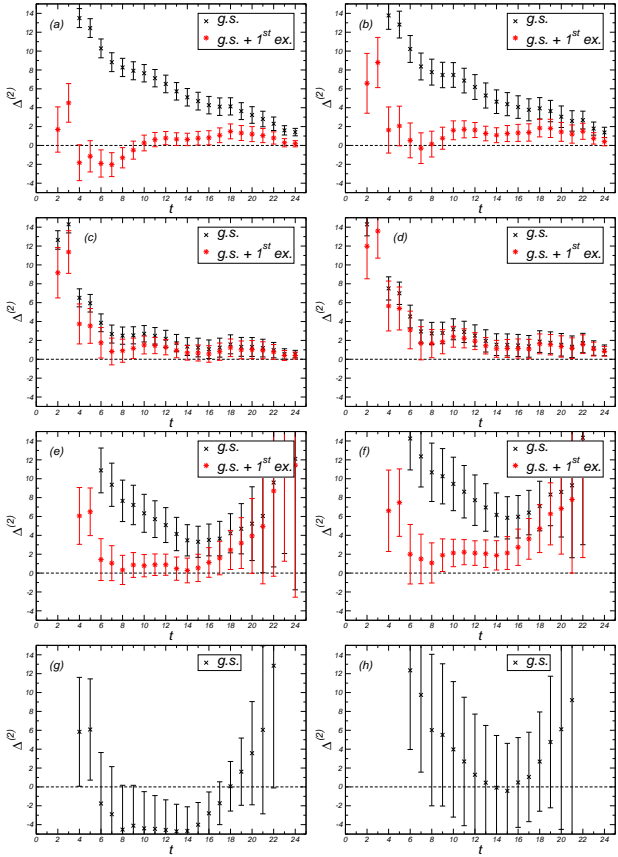

  \centering
      \psfig{width=4.0cm,file= pion.D1500_fit.eps}
     \psfig{width=4.0cm,file= pion_px1_py0_pz0.D1500_fit.eps}
      %\hspace{.3cm}
      \psfig{width=4.0cm,file= sum_rho.D1500_fit.eps}
      \psfig{width=4.0cm,file= sum_rho_px1_py0_pz0.D1500_fit.eps}
      % \hspace{.3cm}
      \psfig{width=4.0cm,file= a0.D1500_fit.eps}
       \psfig{width=4.0cm,file= a0_px1_py0_pz0.D1500_fit.eps}
      % \hspace{.3cm}
      \psfig{width=4.0cm,file= a1.D1500_fit.eps}
      \psfig{width=4.0cm,file= b1.D1500_fit.eps}
  \caption{Data($\sigma=3.6 - \sigma=3.6$)-fit percentage deviation,
    $\Delta^{(2)} \equiv 100 \tfrac{\mathrm{data} -
    \mathrm{fit}}{\mathrm{data}}$. Black points indicate the
    contribution of the ground-state to the fit, the red points
    include both the ground and first excited states. (a)
    $\eta_c(\gamma_5)$, $\vec{p}=(000)$, (b) $\eta_c(\gamma_5)$,
    $\vec{p}=(100)$, (c) $\psi(\gamma_k)$, $\vec{p}=(000)$, (d)
    $\psi(\gamma_k)$, $\vec{p}=(100)$, (e) $\chi_{c0}(1)$,
    $\vec{p}=(000)$, (f) $\chi_{c0}(1)$, $\vec{p}=(100)$, (g)
    $\chi_{c1}(\gamma_k \gamma_5)$, $\vec{p}=(000)$, (h) $h_c(\gamma_k
    \gamma_4)$, $\vec{p}=(000)$}
  \label{fig:two_point_fits}
\end{figure}

Fits are shown in Figure \ref{fig:two_point_fits} and Table \ref{tab:two}. 

\begin{table}[h]
  \centering
  \begin{tabular}{ccccc}
 $\Gamma$   & $J^{PC}$ & & $m^{(0)}$/MeV & $m^{(1)}$/MeV \\
\hline 
$\gamma^5$ & $0^{-+}$ & $\eta_c$ & $2819(4)$ & $3621(17)$ \\
$\gamma^k$ & $1^{--}$ & $\psi$ & $2916(4)$ & $3810(22)$ \\
$1$ & $0^{++}$ & $\chi_{c0}$ & $3287(14)$ & $4168(52)$ \\
$\gamma^5 \gamma^k$ & $1^{++}$ & $\chi_{c1}$ & $3401(28)$ & - \\
$\epsilon_{ijk} \gamma^i \gamma^j$ & $1^{+-}$ & $h_c$ & $3374(35)$ & - 
  \end{tabular}
  \caption{\label{tab:two}Local interpolating fields, $\bar{\psi} \Gamma \psi$ and
    extracted charmonium spectrum, using the determination of the
    lattice spacing as described in the text.}
\end{table}

The ground state masses extracted are observed to be systematically around $5\%$
lighter than the experimental masses - this is a result of imperfect
tuning of the quark mass parameter; for this first study of radiative
transitions we did not attempt a more accurate tuning.

We extract a $J/\psi - \eta_c$ hyperfine splitting of $97(6) \mathrm{MeV}$. This is
probably $10\%$ larger than it would be if we had correctly tuned the
quark mass, assuming the hyperfine splitting scales like $m_c^{-2}$;
with this correction we are in rather good agreement with the result
$88(4) \mathrm{MeV}$ from the closely related overlap fermion action
on an isotropic lattice at Wilson $\beta = 6.3$\cite{Tamhankar:2005zi}. These
two results are significantly larger than typical results using other
improved actions such as clover. Assuming that there is very little
change in scaling to the continuum this leaves a deviation from
experiment of $\sim 30 \mathrm{MeV}$, which is consistent with the
sort of deviations suggested by unquenching\cite{Gottlieb:2005me,
diPierro:2003bu} and inclusion of disconnected diagrams\cite{McNeile:2004wu,
deForcrand:2004ia}.

The excited state masses we are able to extract are clearly too large
in comparison with experiment. The same effect was observed in the
anisotropic clover study of \cite{Okamoto:2001jb} when they set their
scale using $r_0$, and to a lesser extent in the
studies of \cite{Chen:2000ej} and \cite{Liao:2002rj}. One possible
explanation may be that, as discussed above, several excited states
are contributing at intermediate times and that our fitting is not
able to fully resolve them such that we end up with one effective
excited state ``mocking up'' the effect of several\footnote{Consider
in particular the vector channel where the second excited state
$\psi(3770)$ is very near to the first $\psi(3686)$, although its
$^3D_1$ nature may lead to a reduced overlap with the local
interpolating field}. Including a second excited state into the fit
hypothesis did not produce stable fits.

\begin{figure}[h]
  \centering

      \psfig{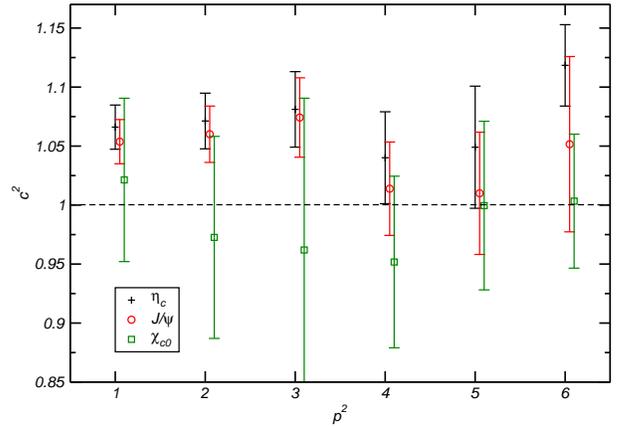}

  \caption{Speed of light extracted from meson dispersion relation: $c^2 \equiv \tfrac{E^2(\vec{p}) -
      m^2}{\vec{p}^2}$ }
  \label{fig:dispersion}
\end{figure}

The multi-correlator fits were performed on the data with many
$\vec{p}$  to determine the dispersion relation for our meson
states. This is particularly relevant on an anisotropic lattice where
parameters in the fermion action must be tuned to match the
gauge sector anisotropy and set the speed of light $c^2 \equiv
\tfrac{E^2(\vec{p}) - m^2}{\vec{p}^2}$ equal to one. The results are
shown in figure \ref{fig:dispersion} where we observe that our tuning
was good to the $3\%$ level on $c=1.032(9)$. 

\subsection{Charmonium decay constants}

Apart from the hyperfine splitting, the only other quantities we
consider that might be significantly distorted by the quenched
approximation are decay constants, which are related to the meson
wavefunction at the origin and are hence sensitive to short-distance
physics. We discussed above %(simulation section) 
that since we set the scale using long-distance dominated quantities,
we expect distortion of short-distance quantities owing to the
incorrect running of the QCD coupling. To be specific, the quenched
coupling at short distances is too small and as such we expect the
wavefunction at the origin to be depleted and hence the decay
constants to be too small with respect to experiment.

\begin{figure}[h]
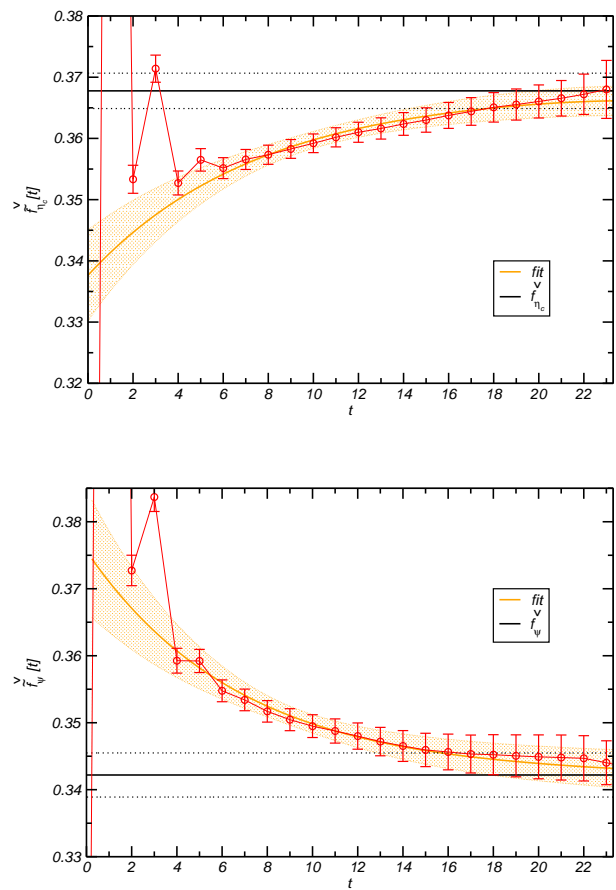

  \centering
      \psfig{width=8cm,file= f.pion.D1500_fit.eps}
      \psfig{width=8cm,file= f.rho.D1500_fit.eps}
  \caption{Data points and fits ($t=9-22$) - see equations
    \protect\eqref{eq:pion_decay}, \protect\eqref{eq:rho_decay}. The black line indicates the ground
    state contribution, $f^{(0)}$. Upper plot - $\eta_c$, lower plot - $\psi$.}
  \label{fig:decay_consts}
\end{figure}

We define the $\eta_c$ and $\psi$ decay constants via\footnote{An
  alternative definition for the vector decay constant is often used
  in lattice simulations}
\begin{align}
  \label{eq:decay}
  \langle 0 | A^\mu  |
  \eta_c(\vec{q})\rangle &=     \langle 0 | \bar{\psi}(0) \gamma^\mu \gamma^5 \psi(0) |
  \eta_c(\vec{q})\rangle = i f_{\eta_c} q^\mu  \nonumber\\
    \langle 0 | V^\mu  |
  \psi(\vec{q})\rangle &=
 \langle 0 | \bar{\psi}(0) \gamma^\mu \psi(0) |
  \psi(\vec{q}, \lambda) \rangle = m_{\psi} f_{\psi}
  \epsilon^\mu(\vec{q}, \lambda). 
\end{align}
In the pseudoscalar ($P = \bar{\psi} \gamma^5 \psi$) case we can form
the following object
\begin{equation}
 \tilde{f}_{\eta_c} \equiv  \frac{2
    Z_A \sum_{\vec{x}} \langle P_{\sigma}(\vec{x},t) A^0 \rangle }{Z^{(0)}_\sigma (e^{- m^{(0)} t} - e^{- m^{(0)} (48 -t)})},\nonumber
\end{equation}
where the subscript $\sigma$ indicates that the interpolating field is
smeared with radius $\sigma$ and where the superscript $(0)$ indicates
the ground state. The renormalisation of the local axial current,
$Z_A$, will be discussed in the next section. Using \eqref{eq:decay},
$\tilde{f}_{\eta_c}$ is equal to
\begin{equation}\label{eq:pion_decay}
  f_{\eta_c^{(0)}} + f_{\eta_c^{(1)}}
  \frac{Z^{(1)}_\sigma }{Z^{(0)}_\sigma } \frac{e^{- m^{(1)} t} - e^{-
      m^{(1)} (48 -t)}   }{e^{- m^{(0)} t} - e^{- m^{(0)} (48 -t)}}  + \hdots
\end{equation}
We fit the lattice data over the range $t=9-22$ with this form using
the $Z^{(0,1)}_\sigma$, $m^{(0,1)}$ found in the earlier two-point
function fits and considering $f_{\eta_c}^{(0,1)}$ to be free fit
parameters - the result for the smearing choice $\sigma=3.6$ is shown
in figure \ref{fig:decay_consts}(a).

We can perform an analogous fit to the vector two-point data,
\begin{align}
 \tilde{f}_{J/\psi} &\equiv  \frac{2
    Z_V \sum_{\vec{x}}\sum_i \langle V^i_\sigma(\vec{x},t) V^i
    \rangle }{3 Z^{(0)}_\sigma (e^{- m^{(0)} t} + e^{- m^{(0)} (48
      -t)})} \nonumber \\
&= f_{\psi^{(0)}} + f_{\psi^{(1)}}
  \frac{Z^{(1)}_\sigma}{Z^{(0)}_\sigma } \frac{e^{- m^{(1)} t} + e^{-
        m^{(1)} (48 -t)}   }{e^{- m^{(0)} t} + e^{- m^{(0)} (48  -t)}}
    + \hdots \label{eq:rho_decay}
\end{align}
the resulting fit with $\sigma=3.6$ being shown in figure
\ref{fig:decay_consts}(b). The renormalisation of the local vector
current, $Z_V$ is discussed in the next section.

Experimentally, the vector decay constants can be extracted from the
leptonic decay widths\footnote{Note that we do not include the
explicit ``QCD correction'' factor $(1- \tfrac{16}{3 \pi} \alpha_s)$
\cite{Ebert:2003rh} under the assumption that all gluon effects, even
hard gluons close to the charm quark mass scale, are included in the
path integral computed},

\begin{equation}
  \Gamma(\psi \to e^+ e^-) = \frac{4 \pi}{3} \frac{4}{9} \alpha^2
  \frac{f_\psi^2}{m_\psi},\nonumber 
\end{equation}
so that, using the PDG averages\cite{PDBook} we have $f_{J/\psi} =
411(7) \mathrm{MeV}$ and $f_{\psi'} = 279(8) \mathrm{MeV}$. The
$\eta_c$ decay constants are rather more difficult to measure; the
only result available comes from $B \to \eta_c K$
\cite{Edwards:2000bb} and involves using a factorisation approximation
to yield $f_{\eta_c} = 335(75) \mathrm{MeV}$. It is clear that within
errors $f_{J/\psi} \approx f_{\eta_c}$ as one would expect from the
non-relativistic quark model in which the $J/\psi$ and the $\eta_c$
differ only by $v/c$ suppressed spin-dependent terms.

From our fits we extract (the scale setting on an anisotropic lattice
is discussed in appendix \ref{scale})
\begin{align}
  |f_{J/\psi}| &= 399(4) \mathrm{MeV}; \quad |f_{\eta_c}| = 429(4)(25) \mathrm{MeV}
 \nonumber \\
 |f_{\psi'}| &= 143(81) \mathrm{MeV}; \quad |f_{\eta'_c} | = 56(21)(3)
 \mathrm{MeV} \nonumber
\end{align}
The first error is statistical and the second, estimated systematic
error, is of order $5\%$ and accounts for the small violations of $Z_V
= Z_A$ we find for our domain-wall fermions, see the following section
for details.

The ground state values are in reasonable agreement with the
experimental extractions, perhaps indicating that even here, with a
short-distance dominated quantity, the quenched approximation is only
impacting at the few percent level. It is important however to recall
that our charm quark mass is a little small, and it is possible that
the discrepancy caused by this is approximately canceling the
quenching discrepancy. A further calculation with a slightly larger
quark mass would allow for an interpolation to the true charm quark
mass and a more precise estimation; since our primary interest is with
radiative transitions, this additional computation has not yet been
carried out.

The excited state decay constants extracted are somewhat smaller than
one might expect, however we recall that we were not very successful
in extracting the experimental masses for these states from two-point
functions. It is possible that the excited state exponential is
absorbing the effect of several excited states - a more systematic
study of interpolating fields with the intent of isolating a field
with large overlap on to the first excited state would clarify the
issue.

\section{Three-point functions}\label{three}

The field theoretic quantity that most simply encodes radiative
transitions is the three-point function:
\begin{multline} \label{threept}
 \Gamma^{(3)}_{f \Gamma i}(\vec{p}_f, \vec{q}; t_f, t) \\
= \sum_{\vec{x},\vec{y}} e^{-i
   \vec{p}_f.\vec{x}}e^{+i \vec{q}.\vec{y}} \langle O_f(\vec{x}, t_f) \;
 \bar{\psi} \Gamma \psi(\vec{y}, t) \; O_i^\dag(\vec{0},0) \rangle 
\end{multline}
The $O_{i,f}$ are interpolating fields chosen to have some
overlap with whatever meson states we are interested in. The current
insertion, $\bar{\psi} \Gamma \psi$, in our case is chosen to
have vector quantum numbers so as to attach to an external photon. If
we opt to use fermion bilinears for the meson interpolating fields,
the possible Wick contractions fall into three classes:
\begin{figure}[t]
  \centering
      \psfig{width=9cm,file= 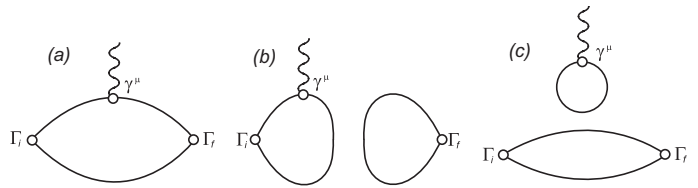}
  \caption{Wick contractions for charmonium three-point functions. (a) connected, (b) OZI-suppressed, (c) disconnected loop}
  \label{fig:wick}
\end{figure}
\begin{itemize}
\item[(a)] the connected diagram of figure \ref{fig:wick}(a);
\item[(b)] disconnected diagrams where an intermediate state without charm
  quarks appears, e.g. figure \ref{fig:wick}(b);
\item[(c)] diagrams in which the photon couples to a closed fermion loop,
  e.g. figure  \ref{fig:wick}(c).
\end{itemize}

Disconnected diagrams of the type in figure \ref{fig:wick}(b) should
be small - in charmonium, OZI suppression is believed to be strong,
which has at least a partial explanation in QCD perturbation theory
since there the intermediate two or three gluon state is suppressed as
a power of $\alpha_s(\mu \sim m_c)$. We do not calculate diagrams of
this type in the current computation.

Diagrams with the photon coupling to a closed fermion loop, as in
figure \ref{fig:wick}(c), are zero in the case of form-factors
($i=f$), on the basis of charge-conjugation invariance of the QCD
action\cite{Draper:1988bp}. In the case of transitions ($i \neq f$),
while the ``$C$-odd'' connected two-point function vanishes in the
ensemble average, the inclusion of the ``$C$-odd'' disconnected loop acts to
compensate leaving a non-zero quantity. In the light quark sector such
a diagram would be required to allow $\omega$ or $\phi$ VMD-style
processes, although such disconnected contributions to the QCD vector
correlator two-point functions are believed small as evinced by the
small empirical $\omega-\phi$ mixing, and hence are unlikely to
contribute much to the transitions under study. In addition, in the particular
case that the vector current is electromagnetic, we also
have a suppression of this disconnected contribution owing to the
sum of $u,d,s$-quark charges being zero and there being an approximate
$SU(3)$ flavour symmetry. In our quenched charmonium
calculation we might consider allowing a charm-quark loop, but we
expect the contribution to be small dynamically owing to the large
charm quark mass and we do not explicitly include it in our
computation.

We are left with the connected diagrams, figure \ref{fig:wick}(a),
with the possibility of light quark loops removed since we are working
in the quenched approximation. We compute these diagrams on the
lattice using the sequential source
technique\cite{Bonnet:2004fr}. This is simply a lattice technology
that enables computation of the required propagators with only two inversions of the Dirac matrix and
without computing an all-to-all propagator. The connected three-point
function written in terms of fermion propagators
$G^{\alpha\beta}_{ij}(x,y)= \psi^\alpha_i(x) \bar{\psi}^\beta_j(y)$ is
\begin{multline} 
\Gamma^{(3)}_{f\Gamma i}(\vec{p}_f, \vec{q}; t_f, t ) \nonumber \\
= - \sum_{\vec{x},
\vec{y}} e^{-i\pfdotx}
e^{+i\vec{q}\cdot\vec{y}} \;   \mathrm{tr} \Big\langle G(0,x)\, \Gamma_f\, G(x,y)\, \Gamma\, G(y,0)\, \Gamma_i  \Big\rangle.
\end{multline}
In the sequential source technique, this is factorised as
\begin{multline}
\Gamma^{(3)}_{f\Gamma i}(\vec{p}_f, \vec{q}; t_f, t ) \nonumber \\
= - \sum_{\vec{y}} e^{+i\vec{q}\cdot\vec{y}} \;   \mathrm{tr} \Big\langle \gamma_5 H_{\Gamma_f}^\dag(y, 0; t_f; \vec{p}_f) \gamma_5  \, \Gamma\, G(y,0)\, \Gamma_i \gamma_5  \Big\rangle ,
\end{multline}
with
\begin{equation}
H_{\Gamma_f}(y,0; t_f; \vec{p}_f) \equiv \sum_{\vec{x}} e^{i \pfdotx} G(y,x) \gamma_5 \Gamma_f^\dag \gamma_5 G(x,0) \gamma_5.\nonumber
\end{equation}
If the propagator $G(y,0)$ is computed via an inversion of the Dirac
matrix, $M^{ik}_{\alpha\gamma}(x,z) G^{kj}_{\gamma\beta}(z,y) = \delta_{ij}
\delta_{\alpha \beta} \delta_{xy}$,  then one can obtain
$H_{\Gamma_f}(y,0; t_f; \vec{p}_f)$ via one further inversion using
$G(y,0)$ in the inversion source, 
\begin{equation}
M(z,y) H_{\Gamma_f}(y,0; t_f; \vec{p}_f) = \delta_{t_z, t_f} e^{i
  \pfdotz} \gamma_5 \Gamma_f^\dag \gamma_5 G(z,0) \gamma_5.\nonumber
\end{equation}
The cost is that we have specified the momentum and species of the
sink particle and thus we must perform a new calculation for each new
particle or momentum. For this reason we performed simulations only
with $\vec{p}_f=(000)$ and $(100)$. We are able, however, in this
formalism to insert any current, $\Gamma$, with any momentum,
$\vec{q}$. Our particular interest is with the vector insertions
$\Gamma = \gamma^\mu$. In addition, we can insert any meson source
gamma matrix structure $\Gamma_i$ without further inversions.

The three-point function can be related to meson matrix elements of
the vector current by inserting complete sets of states into
\eqref{threept} yielding
\begin{widetext}
\begin{equation}\label{}
 \Gamma^{(3)}_{f \Gamma i}(\vec{p}_f, \vec{q}; t_f, t)  = \sum_{f, i}  \frac{e^{- E_f t_f}
     e^{- (E_i - E_f) t } }{2 E_f(\vec{p}_f)\;2 E_i(\vec{p}_i) } \langle 0 |
   \bar{\psi}\Gamma_f \psi(\vec{0}, 0) |f(\vec{p}_f, r_f)\rangle \langle f(\vec{p}_f, r_f) |
   \bar{\psi} \Gamma \psi(\vec{0}, 0)   |i(\vec{p}_i, r_i) \rangle\Big( \langle 0 |  \bar{\psi}\Gamma_i \psi(\vec{0},0) |  i(\vec{p}_i, r_i) \rangle \Big)^*, \nonumber
\end{equation}
\end{widetext}
so that (in Euclidean space) we have a sum of
decaying exponentials. If $t$ is sufficiently far from $0$ and
$t_f$ the excited state contributions should have decayed away
leaving us with only the ground state transition matrix element. The
removal of the excited state contributions is assisted by using
the interpolating field smearing
choices we made to improve plateaux in the two-point functions.

In our computation we place the source at $t=0$ and the sink at
$t=24$ and consider insertions of the vector current on all
time-slices. Since we have an antiperiodic lattice of length $48$, this makes
the second half of the lattice symmetric with the first and in
practice we average the two halves to get our form-factor results,
thus improving the statistics.

The energies of meson states and the overlap of the interpolating
fields with given states ($Z$) are extracted from the meson two-point
functions as described in section \ref{spectrum}. This allows us to
adopt a fitting method for extraction of matrix elements from the
three-point functions. The procedure involves writing the three-point
function on a given time-slice as the sum of products of a (known)
propagation factor $P = \tfrac{Z_f Z_i}{4 E_f E_i} e^{- E_f t_f}
e^{-(E_i - E_f) t }$, a (known) kinematic factor (e.g. $K^\mu = (p_f +
p_i)^\mu$ for the $\eta_c \gamma^\mu \eta_c$ case) and an unknown
form-factor, i.e.
\begin{equation}
  \Gamma(p_f, p_i; t) = \sum_n P(p_f, p_i; t) \cdot K_n(p_f, p_i)
  \cdot f_n(Q^2). \nonumber
\end{equation}
One can then form a vector of three-point functions that all have the
same $Q^2 = |\vec{p}_f - \vec{p}_i|^2 - (E_f - E_i)^2$
(i.e. rotationally equivalent momentum combinations and different
combinations of Lorentz indices, labeled here by $a,b,c \ldots$) and a
matrix with entries $P K$, to define a linear system
\begin{equation}
  \label{eq:svd}
  \begin{bmatrix} \Gamma(a; t)\\ \Gamma(b;t ) \\ \Gamma(c; t)\\ \vdots
  \end{bmatrix} = \begin{bmatrix} P(a; t) K_1(a) & P(a; t) K_2(a) &
  \ldots \\ P(b; t) K_1(b) & P(b; t) K_2(b) \\ P(c; t) K_1(c)
& P(c; t) K_2(c) \\
\vdots & & \ddots \end{bmatrix} \begin{bmatrix}f_1(Q^2)[t] \\
f_2(Q^2)[t] \\ \vdots \end{bmatrix}, \nonumber
\end{equation}
which we can invert with SVD to find the form-factors $f_n(Q^2)[t]$.

Single elimination jackknife is implemented such that statistical
errors are propagated through the inversion process. This proceeds by
eliminating one configuration and averaging the remaining
three-point function ensemble for use in the right-hand-side above and
computing the propagation and kinematic factors using the equivalent
``single-elimination bin'' from the $Z,E$ ensemble extracted from the
two-point function fits. The SVD inversion is then performed for this ``single-elimination bin''. This is repeated for
elimination of each
configuration yielding a configuration
ensemble of solutions $\{ f_n \}$.

\begin{figure}[h]
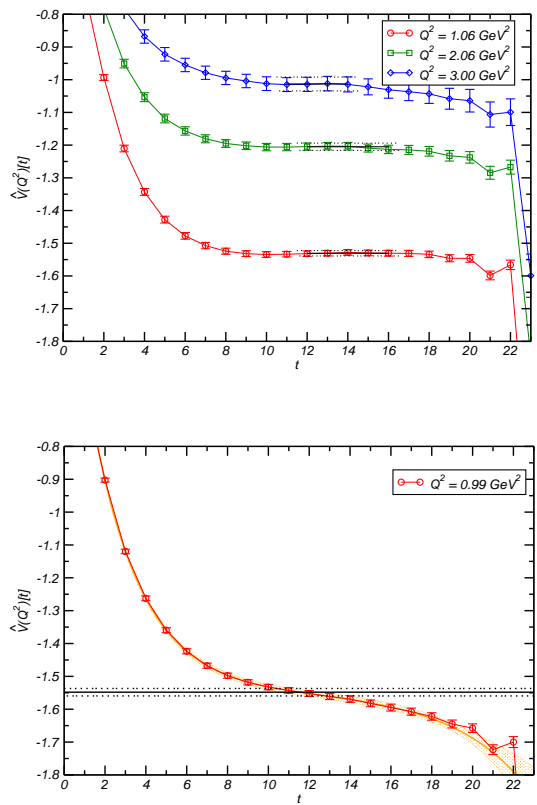

  \centering
      \psfig{width=7.0cm,file= pion_rho_fit_fold.out0_s.eps}

 \psfig{width=7.0cm,file= rho_pion_fit_fold.out0_qsq_0.987104_fit.eps}
  \caption{Examples taken from $J/\psi \to \eta_c \gamma$ (a) plateau
    fits (b) fit by the function \protect\eqref{eq:fit_correlator}, parameters are
    $\hat{V}(Q^2) = -1.55(1)$, $\mathfrak{f}_i = 1.45(3)$,
    $\mathfrak{f}_f = -0.42(14)$, $\mathfrak{m}_i =0.41(1)$, $\mathfrak{m}_f =0.27(7)$ }
  \label{fig:fit_correlator}
\end{figure}

The form-factors $f_n(Q^2)[t]$ should have no $t$ dependence once excited states have
decayed away - i.e. they have plateaux. This linear inversion formalism can also
include excited state contributions explicitly, provided their
energies and $Z$'s have been extracted from the two-point functions -
we would simply enlarge the space of the linear system. In the present
analysis we consider only ground state contributions - our poor
estimation of the mass of excited states leads us to suspect that we
are not able to isolate purely the first excited state. In a few cases
we will find that an $f_n(Q^2)[t]$ does not reach a plateau within the
available time, in such cases we fit in a region around $t=12$ (away
from the domain-wall oscillations near $t=0,24$), with a form
\begin{equation}
\label{eq:fit_correlator} 
 f_n(Q^2)[t] = f_n(Q^2) + \mathfrak{f}_i e^{- \mathfrak{m}_i t} + \mathfrak{f}_f e^{- \mathfrak{m}_f
    (24-t) },
\end{equation}
with $ f_n(Q^2), \mathfrak{f}_i, \mathfrak{f}_f, \mathfrak{m}_i,
\mathfrak{m}_f$ as variables. A typical result of this fitting is
shown in figure \ref{fig:fit_correlator}(b). Our lack of control over
the contribution of excited states is demonstrated by the extracted
masses, $\mathfrak{m}_{i,f}$, which are typically considerably larger
than expected on the basis of either the true or our lattice first
excited state masses. These fits are rather successful, suggesting
that the single exponential pollution hypothesis is reasonable even
though we expect contributions from several excited states - a
possible explanation might arise if alternating excited state
contributions oscillated in sign (either from the $Z$-factors or from
the excited state transition matrix elements)\cite{Holl:2004un}. In
the following sections, $f(Q^2)$ points extracted in this manner are
displayed in figures by ghosted shapes while points with a clear
plateau appear with solid lines.

A possible explanation for the large values of $\mathfrak{m}_i$ we observe comes from an alternate time-ordering allowed by our anti-periodic lattice. The periodicity implies that there is an image
 of the sink at $t = -24$, such that with the ``current'' insertion at
 timeslice $t>0$ there is a process: creation of sink particle with
 $\Gamma_f$ at $t=-24$, insertion of a ``current'' $\Gamma_i$ at $t=0$ and
 annihilation of a vector particle by $\gamma^\mu$ at $t$. This process
 has time dependence $e^{-24 E_f} e^{- E_\psi(\vec{q})t }$, which
 would correspond in the $\psi - \eta_c$ case to $\mathfrak{m}_i =
 E_\psi(\vec{q}) \sim 0.5$ in lattice units. A limited computation of
 three-point functions with Dirichlet boundary conditions appears to
 confirm this hypothesis.

In practice we can access a number of $Q^2$ values through projecting
various momenta at the sink and the insertion. At the sink we are
limited by the sequential source technique and only use
$\vec{p}_f=(000), (100)$, but at the insertion we project many
momenta, corresponding to $\vec{p}_i$ up to around $(211)$. Higher
momenta were calculated but the signals were increasingly noisy. We
used smearing at both source and sink with $\sigma = 3.6$.

\subsection{Current renormalisation}\label{ZV}  
\begin{figure}[h]
  \centering
      \psfig{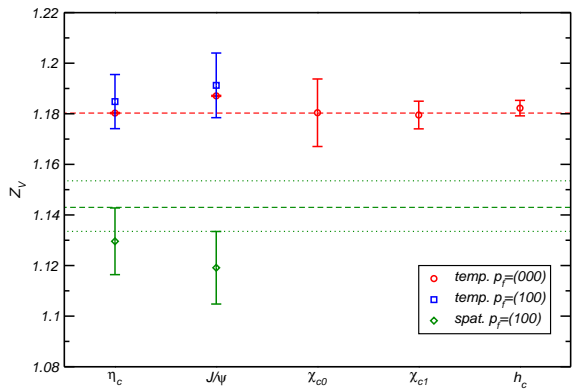}
  \caption{$Z_V$ extracted from two-point/three-point function
    ratios. Green dashed line shows $Z_V^{(0)}/c$. The
  extracted spatial data are obviously compatible with this.   }
  \label{fig:ZV}
\end{figure}

We utilise the simple local vector current, $\bar{\psi}(x) \gamma^\mu
\psi(x)$, which is not conserved with a discretised fermion action and
as such gets multiplicatively renormalised by a factor $Z_V(a)$. We
extract $Z_V$ using a ratio of two-point and three-point functions
evaluated at $Q^2=0$, which in the case of a temporal vector current
is not polluted by excited state transitions\cite{Barad:1984px}. For
the $\eta_c$ and the $\chi_{c0}$ we use
\begin{equation}
  Z^{(\mu)}_V(t) = \frac{p^\mu}{E(\vec{p})} \frac{\tfrac{1}{2}
    \Gamma^{(2)}_{\eta_c \eta_c}(\vec{p}; t_f=24)
  }{\Gamma^{(3)}_{\eta_c \gamma^\mu \eta_c}(
    \vec{p}_f=\vec{p_i}=\vec{p}; t_f=24, t )}\nonumber
\end{equation}
where the factor of $\tfrac{1}{2}$ accounts for the equal contribution
to the two-point function of the source at timeslice $0$ and the image
of the source at timeslice $48$ (recall that we have a temporally
anti-periodic lattice). For the spin-$1$ particles ($J/\psi,
\chi_{c1}, h_c$) we use
\begin{equation}
  Z^{(\mu)}_V(t) = \frac{p^\mu}{E(\vec{p})} \frac{\tfrac{1}{2} \sum_k
    \Gamma^{(2)}_{\psi_k \psi_k}(\vec{p}; t_f=24) }{\sum_k
    \Gamma^{(3)}_{\psi_k \gamma^\mu \psi_k}(
    \vec{p}_f=\vec{p_i}=\vec{p}; t_f=24, t)}\nonumber
\end{equation} 
Since we are working on an anisotropic lattice the renormalisation of
the spatial and temporal components of the vector current could, in
principle, differ. The extracted $Z_V$ are shown in figure
\ref{fig:ZV} where it is clear that there is no significant dependence
upon the particle used in the extraction. The $\eta_c$ extraction is
cleanest and gives:
\begin{equation}
  Z_V^{(0)} = 1.1803(2);  \quad Z_V^{(k)} = 1.130(13).\nonumber
\end{equation}
Note that these numbers differ by a factor comparable to the $c$
values extracted in the dispersion relation analysis, indicating that
wherever a three-momentum $\vec{p}$ appears we should in fact insert
an explicit $c = 1.032$ to account for the imperfect dispersion. In
what follows we will adopt this prescription for the momentum and use
the temporal $Z_V$.

The renormalization factor, $Z_A$, for the four dimensional axial
current operator $A_\mu = \bar{\psi} \gamma_\mu \gamma_5 \psi $ is calculated
using the five dimensional conserved axial current for domain wall
fermions ${\cal A}_\mu$ by the relation~\cite{Blum:2000kn} 
$\langle {\cal A}_\mu(t) A_\mu(0)\rangle = Z_A \langle A_\mu(t)
A_\mu(0)\rangle$. In the limit of $L_5\to\infty$, $Z_V=Z_A$. However,
at finite $L_5$ there is a breaking of chiral symmetry that
is characterized by a additive (and positive) quark mass shift 
called the residual mass $m_{\mathrm{res}}$. The difference of $Z_V$ from
$Z_A$ is proportional to this residual mass. The latter we find to be
consistent with isotropic Wilson gauge calculations at the same spatial
lattice spacing~\cite{Blum:2000kn}.

\section{Charmonium vector form-factors}\label{ff}

Charmonia, unlike charged or flavoured mesons, do not have radiative
form-factors owing to them being eigenstates of charge-conjugation
invariance. At the quark level this comes about through the photon
coupling to both the quark and the anti-quark with equal strength. In
our lattice computation we insert the vector current only on the quark
line and are thus able to access its distribution within the
meson. This can be compared to models.

\subsection{$\eta_c$ ``form-factor''}
\begin{figure}[h]
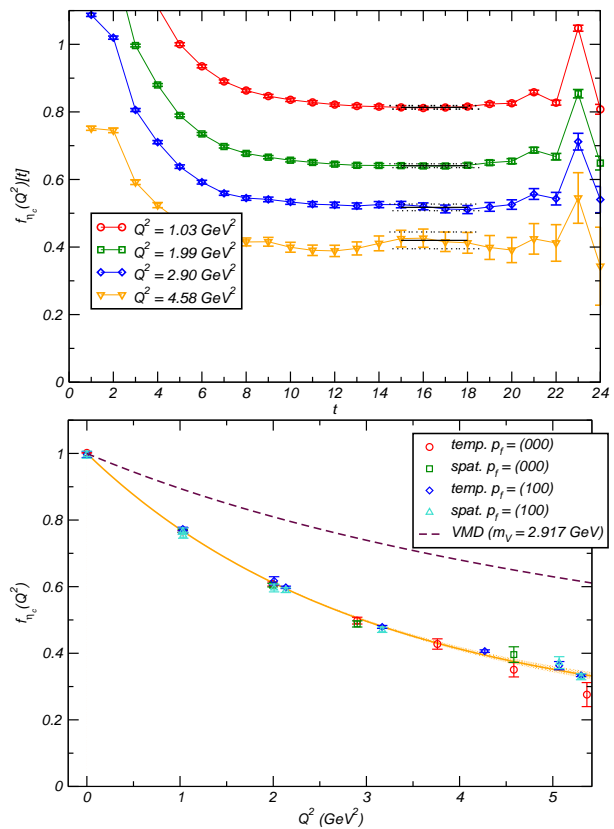

  \centering
      \psfig{width=8.0cm,file= pion_ff_fold.out0_s.eps}
      \hspace{1cm}
      \psfig{width=8.0cm,file= pion_ff_fit.eps}
    
  \caption{$\eta_c$ form-factor (a) typical lattice signal with plateau fits (b) form-factor}
  \label{fig:pion_ff}
\end{figure}

This is defined by the matrix element decomposition
\begin{equation}
  \label{eq:pion_ff}
  \langle \eta_c(\vec{p}_f) | j^\mu(0) | \eta_c(\vec{p}_i) \rangle =
  f(Q^2) (p_i+p_f)^\mu.
\end{equation}
In figure \ref{fig:pion_ff}(a) we display some typical $f(Q^2)[t]$ as
extracted from our simulation in the manner described in the previous
section. We see clear plateau behaviour away from the source and sink
points which is fitted and a value of $f(Q^2)$ obtained. These points
are displayed in figure \ref{fig:pion_ff}(b). Also shown is the
expectation of VMD using the $J/\psi$ mass found in section
\ref{spectrum}, this is seen to give a very poor description of the
data in contrast to what is found in the light-quark sector for the
pion form-factor \cite{Bonnet:2004fr}. This is to be expected on the
basis of a dispersion equation approach - in the light quark sector,
the $\rho$ pole is the nearest left-hand vector singularity in the
complex $Q^2$ plane and the next nearest (neglecting multi-pion cuts
and isospin suppressed poles) is the excited $\rho'(1450)$ which
is relatively distant. In charmonium the nearest pole is the
$J/\psi(3097)$ and the next nearest the $\psi'(3686)$ which is not
sufficiently distant as to be negligible. Hence in charmonium it is
likely that one needs to sum many vector meson poles (of unknown
residue) to agree with $Q^2>0$ data.

We take the more pragmatic approach of fitting the ($Q^2 >0$) data to a
simple analytic form, $\exp \left[-\tfrac{Q^2}{16 \beta^2}\left(1+
    \alpha Q^2\right)\right]$. This fit is also shown in figure \ref{fig:pion_ff}(b)
where one can see it faithfully reproduces the data over the entire
$Q^2$ range considered. The fit parameters so obtained are
\begin{equation}
  \label{eq:pion_beta}
  \beta = 480(3) \mathrm{MeV}; \quad \alpha = -0.046(1) \mathrm{GeV}^{-2}.\nonumber
\end{equation}
One can define a squared ``charge'' radius by $\langle r^2 \rangle = -6
\tfrac{d}{dQ^2} f(Q^2)\big|_{Q^2=0} = \tfrac{6}{16 \beta^2}$, which
yields $\sqrt{\langle r^2 \rangle} = 0.255(2) \mathrm{fm}$, giving some a posteriori justification for our lattice size being
only $\sim 1.2 \mathrm{fm}$.

As mentioned in the introduction, most charmonium phenomenology is
done within the framework of quark-potential models. These models are
usually of the non-relativistic Schr\"odinger equation type, utilising
some phenomenologically justified static potential and incorporating
relativistic corrections to this, such as spin-spin and spin-orbit
terms. By and large these models are successful so it makes sense for
us to compare our lattice results with them in cases where there is
not experimental data.

In the simplest forms of these models we have only Galilean invariance
and not the full Lorentz invariance of a relativistic theory. This is
manifested in the factorisation of the spatial wavefunction of a
charmonium state into centre-of-mass and internal wavefunctions:
\begin{equation}
  |\eta_c(\vec{p})\rangle \propto \int d^3 \vec{R} \,e^{i
    \vec{p}.\vec{R}} \int d^3 \vec{r}\, \psi(\vec{r}),\nonumber
\end{equation}
with $\psi(\vec{r})$ independent of the $\eta_c$ momentum $\vec{p}$,
obtained by solving a Schr\"odinger equation in the rest frame of the
$\eta_c$. This poses problems of frame-dependence when one calculates
matrix elements. Consider calculating $f(Q^2)$ using equation
\eqref{eq:pion_ff} within this model. We want to work in a frame near
to the one in which the $\eta_c$ wavefunction was calculated, to
minimise the neglected boost effects, two choices strike us:
\begin{itemize}
\item {\em the rest frame of the initial $\eta_c$} ($\vec{p}_i=0$,
  $\vec{p}_f=\vec{q}$). Here only the end-state $\eta_c$ could suffer
  boost distortion. We have 
  \begin{equation}
    f\Big(Q^2 = 2m_{\eta_c}^2\big[-1 + (1+
      \tfrac{|\vec{q}|^2}{m_{\eta_c}^2})^{1/2} \big]\Big) \propto \int
    d^3\vec{r}\, \psi^*(\vec{r}) j_0\big(\tfrac{|\vec{q}|r}{2}\big) \psi(\vec{r}).\nonumber
  \end{equation}
  With approximate harmonic oscillator wavefunctions ($\psi(\vec{r})
  \propto e^{-\beta^2 r^2}$) we obtain
  \begin{equation}
    f_{\mathrm{RF}}(Q^2) \to \exp \left[- \frac{|\vec{q}|^2}{16
        \beta^2}\right] =  \exp \left[ - \frac{Q^2}{16 \beta^2}\big( 1
      + \tfrac{Q^2}{4 m_{\eta_c}^2}
      \big) \right]\nonumber
  \end{equation}
\item {\em the Breit frame }($\vec{p}_i = -\tfrac{\vec{q}}{2}$, $\vec{p}_f =
  \tfrac{\vec{q}}{2}$). Here we share the possible boost distortion
  equally between the initial and final states. The quark-model
  evaluation of the matrix element will be essentially identical to the
  previous case with the only difference being that now
  $Q^2=|\vec{q}|^2$. Hence
   \begin{equation}
    f_{\mathrm{BF}}(Q^2) \to \exp \left[- \frac{|\vec{q}|^2}{16
        \beta^2}\right] =  \exp \left[ - \frac{Q^2}{16 \beta^2} \right]\nonumber
  \end{equation} 
and we explicitly see the frame-dependence of the simplest quark
models ($f_{\mathrm{BF}}(Q^2) \neq f_{\mathrm{RF}}(Q^2)$). 
\end{itemize}
If $Q^2$ is
small, all frames will agree, and it is really only here that we have
a unique quark model form. In \cite{Kelly:2002if}, a thorough analysis
of these concerns is applied to the case of the proton form-factors.

We note in passing that light-front quark models have the same sort of
problems - although boosts along a given axis become simple
(kinematical), which may admit consistency up to higher $Q^2$, we lose
rotational invariance, and this can lead to extra form-factors not
present in a picture with $O(3)$ symmetry\cite{Carbonell:1998rj}.

In the quark model of ISGW \cite{Isgur:1988gb}, form-factors have a
dependence near $Q^2=0$ of the form $\exp \big[-\frac{Q^2}{16
\beta^2 \kappa^2}\big]$, with $\kappa \sim 0.7$. The $\kappa$
factor was added by hand to better describe the pion form-factor and
certain heavy quark transitions and was ascribed an origin in
relativistic corrections\footnote{Comparing a calculation using full
Dirac spinors with the more typically applied non-relativistic spinors
one obtains just such a suppression \protect \cite{Eric-private} }. An
alternative origin might be the effect of gluonic degrees-of-freedom
not considered in the simple potential model\cite{thesis,
Isgur:1999kx}. In \cite{Close:2005se} a Schr\"odinger equation with a
Coulomb plus linear potential and spin-dependent corrections was
solved using a variational harmonic oscillator basis - they found for
the $\eta_c$ that $\beta = 710 \mathrm{MeV}$. Hence $\kappa
\beta = 500 \mathrm{MeV}$ which is in rather good agreement with
our $\beta = 480(3) \mathrm{MeV}$, suggesting that the potential
model, with the relativistic correction, is capturing at least some of
the $Q^2 \approx 0$ physics we have on the lattice.

\subsection{$J/\psi$ ``form-factors''}
\begin{figure}[t]
  \centering
 
      \psfig{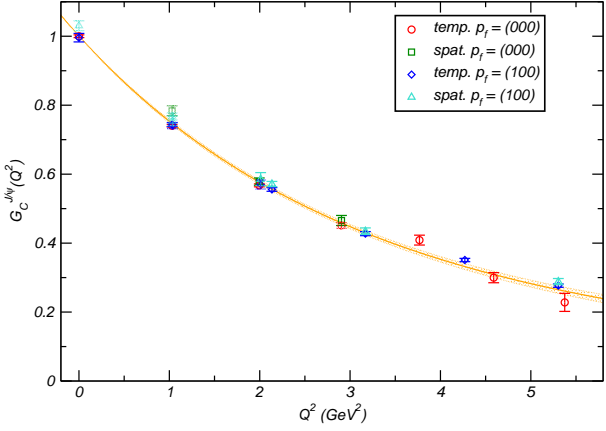}
    
  \caption{$J/\psi$ ``charge'' form-factor}
  \label{fig:rho_c_ff}
\end{figure}

A vector particle has three form-factors, a suitable set being those
of definite multipolarity: charge - $G_C(Q^2)$, magnetic dipole -
$G_M(Q^2)$ and quadrupole - $G_Q(Q^2)$. These are defined in terms of
the standard Lorentz covariant decomposition
\begin{widetext}
\begin{align}
  \langle V(\vec{p}_f, r_f) | j^\mu(0) | V(\vec{p}_i, r_i) \rangle &=
  - (p_f+p_i)^\mu \Big[ G_1(Q^2) \epsilon^*(\vec{p}_f, r_f)\cdot
  \epsilon(\vec{p}_i, r_i) + \tfrac{G_3(Q^2)}{2 m_V^2}
  \epsilon^*(\vec{p}_f, r_i).p_i \, \epsilon(\vec{p}_i, r_i).p_f \Big]\nonumber \\
&\quad+ G_2(Q^2) \Big[ \epsilon^\mu(\vec{p}_i, r_i) \epsilon^*(\vec{p}_f, r_f).p_i
+  \epsilon^{\mu*}(\vec{p}_f, r_f) \epsilon(\vec{p}_i, r_i).p_f
\Big] \nonumber
\end{align}
\end{widetext}
by
\begin{align}
  G_C &= \left(1+\tfrac{2}{3}\eta \right) G_1 - \tfrac{2}{3}\eta G_2 +
  \tfrac{2}{3}\eta (1+\eta) G_3 \nonumber\\
G_M &= - G_2 \nonumber\\
 G_Q &= G_1 - G_2 + (1+\eta) G_3 ,\nonumber
\end{align}
with $\eta = \tfrac{Q^2}{4m^2}$. In figure \ref{fig:rho_c_ff} we
display our lattice points and a fit to them. The charge multipole is
fitted by the same function we used for the $\eta_c$ with resulting
parameters:
\begin{equation}
  \label{eq:rho_beta}
  \beta_C = 470(7) \mathrm{MeV}; \quad \alpha = -0.022(7) \mathrm{GeV}^{-2}.
\end{equation}
This similarity of these parameters to those extracted for the
$\eta_c$ indicate that these two particles have spatial wavefunctions
that are rather alike.
\begin{figure}[h]
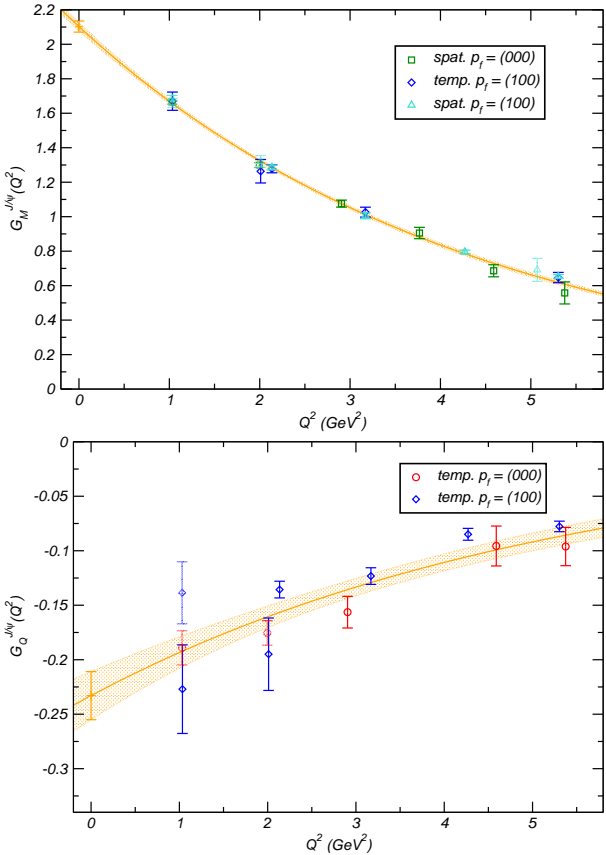

  \centering
      \psfig{width=8.0cm,file= rho_M1_fit.eps}
      %\hspace{1cm}
      \psfig{width=8.0cm,file= rho_C2_fit.eps}
    
  \caption{$J/\psi$ form-factors (a) magnetic (b) quadrupole}
  \label{fig:rho_mq_ff}
\end{figure}
The magnetic dipole and quadrupole form-factors, displayed in figure
\ref{fig:rho_mq_ff}, have fits of the form $G(0) \exp -\tfrac{Q^2}{16
\beta^2}$ which are seen to be successful and yield
\begin{align}
  \label{eq:rho_beta2}
  \beta_M &= 520(8) \mathrm{MeV}; \quad G_M(0) = 2.10(3) \nonumber
  \\
\beta_Q &= 580(44) \mathrm{MeV}; \quad G_Q(0) = - 0.23(2).\nonumber
\end{align}
The quadrupole form-factor at $Q^2=0$ gives us the quadrupole moment
of the $J/\psi$ and, via a quark-model interpretation, access to the
quark-antiquark $D$-wave admixture in the dominantly $S$-wave
$J/\psi$. The quadrupole moment operator is \cite{LandLifs}
$\mathsf{Q} = r^2 ( 3 \cos^2 \theta -1) = \sqrt{\tfrac{16 \pi}{5} }
r^2 Y^0_2(\theta, \phi)$. Then it is clear that $\langle S |
\mathsf{Q} | S \rangle = 0$, while $\langle S | \mathsf{Q} | D \rangle
\neq 0$. In the approximate harmonic oscillator basis, we have
\begin{equation}
  \langle J/\psi | \mathsf{Q}  | J/\psi \rangle = 2\sqrt{3} a_S
  a_D \beta^{-2},\nonumber
\end{equation}
where $a_L$ is the $L$-wave amplitude in the $J/\psi$
wavefunction. Since $G_Q(0) = m_{J/\psi}^2 \langle \mathsf{Q}
\rangle$, using our lattice $J/\psi$ mass we have $\langle \mathsf{Q}
\rangle = - 0.027(3)  \mathrm{GeV}^{-2}$. With the extracted quadrupole
$\beta_Q$ value, and assuming $a_S \approx 1$ we find $a_D = - 2.6(5)
\times 10^{-3}$. This very small value can be used to constrain the
size of any tensor term one might wish to add to a charmonium model
Hamiltonian. We remind the reader that any contribution from coupled-channel $D\bar{D}$ loops will not be present in our quenched computation.

The magnetic dipole moment, $\mu_{J/\psi}$, can be extracted using $G_M(0) = 2
M_{J/\psi} \;\mu_{J/\psi}$ - in units of the $J/\psi$ mass it is
$1.05(2)$. Within the simple non-relativistic quark model this quantity can be
expressed as
\begin{equation}
  \label{eq:mag_mom}
  (1+\delta)\left[ (1+\kappa_c)|a_S|^2 + \tfrac{1}{4}(1- 2\kappa_c)
    |a_D|^2 \right],\nonumber
\end{equation}
where $\delta = \tfrac{M_{J/\psi}-2m_c}{2m_c}$ and $\kappa_c$ is the
anomalous magnetic moment of the charm quark. With the $D$-wave
admixture extracted from the quadrupole moment, the second term in
square brackets is negligible. If one had a way of unambiguously
setting the charm quark mass one could use this expression to extract
the charm quark anomalous magnetic moment - however this is not
possible even in principle - the quark mass is a renormalisation
scheme dependent quantity. In quark potential models it is usually
tuned along with other parameters to give a good description of the
spectrum. A conservative interpretation of our extracted value would
be that it could be explained in a model with a small or zero
anomalous quark moment and a charm quark ``mass'' slightly less than
half the $J/\psi$ mass. This would be interesting application of the
EFT method of pNRQCD \cite{Brambilla:2005zw}- within this picture one
has, at a given order in the power counting, a relationship between
the charm quark mass and the $J/\psi$ mass in terms only of the
(determined) strong coupling, as such one can determine, at the same
order in the power counting, the anomalous moment $\kappa_c$.

\subsection{$\chi_{c0}$ ``form-factor''}

\begin{figure}[h]
  \centering

      \psfig{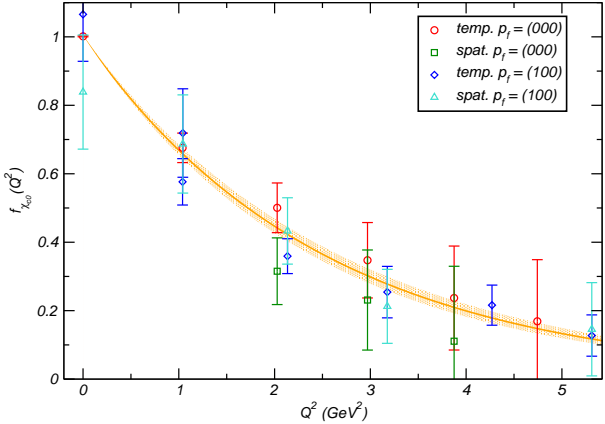}
  \caption{$\chi_{c0}$ form-factor}
  \label{fig:a0_ff}
\end{figure}

This has the same decomposition as does the $\eta_c$, eqn
\eqref{eq:pion_ff}. Our lattice three-point functions here are much
noisier than for the $\eta_c$ and $J/\psi$, but there are sufficiently
clear plateaux to extract the values plotted in figure
\ref{fig:a0_ff}. The data is fitted with the form $f(0)
e^{-\tfrac{Q^2}{16 \beta^2}}$ giving $f(0) = 1.0015(15)$,
consistent with $1$ as it should be, and $\beta = 393(12)
\mathrm{MeV}$, which is roughly consistent with the quark model
value\cite{Close:2005se} $\kappa \beta = 340 \mathrm{MeV}$. That
this is smaller than the $\eta_c, J/\psi$ $\beta$ is simply a
reflection of the fact that the $\chi_{c0}$ is spatially larger due to
the centripetal $P$-wave barrier between the quark and the anti-quark.

\section{Radiative transitions}\label{trans}

\subsection{$J/\psi \to \eta_c \gamma_{M1}$ }

The Minkowski space-time matrix element for this transition can be expressed in terms of
a product of one Lorentz invariant form-factor and one Lorentz
covariant kinematic factor:
\begin{equation}
  \label{eq:rhopi}
  \langle \eta_c(\vec{p}\,') | j^\mu(0) | J/\psi(\vec{p}, r)\rangle =
  \frac{2 V(Q^2)}{m_{\eta_c} + m_{\psi}}
  \epsilon^{\mu\alpha\beta\gamma} p'_\alpha p_\beta
  \epsilon_\gamma(\vec{p}, r).
\end{equation}
This decomposition is parity invariant and, with $V(Q^2)$ real,
time-reversal invariant. The Lorentz invariant matrix element for
$J/\psi \to \eta_c \gamma^*(Q^2)$ is the contraction of
\eqref{eq:rhopi} with a final-state photon polarization vector:
\begin{equation}
  {\cal M}_{r,r_\gamma} = \epsilon^*_\mu(\vec{q}, r_\gamma)  \langle \eta_c(\vec{p}') | j^\mu(0) | J/\psi(\vec{p}, r)\rangle.\nonumber
\end{equation}
The decay width with an on-shell photon is
\begin{equation}
  \label{eq:width_rhopi}
  \Gamma(J/\psi \to \eta_c \gamma) = \int d\Omega_{\hat{q}}
  \frac{1}{32\pi^2}\frac{|\vec{q}|}{m_\psi^2} \frac{1}{3} \sum_{r,
    r_\gamma} \big|{\cal M}_{r,r_\gamma} \big|^2,
\end{equation}
which contains a sum over the final state photon polarisation and an
average over the initial $\psi$ polarisation. Explicitly then we have
\begin{equation}
  \label{eq:width_rhopi2}
  \Gamma(J/\psi \to \eta_c \gamma) =
  \frac{1}{4\pi}\frac{|\vec{q}|^3}{(m_{\eta_c} + m_\psi)^2}
  \frac{4}{3} \big|V(0) \big|^2.\nonumber
\end{equation}
In our lattice computation we couple only to the quark and not to the
anti-quark and do not include the quark electric charge factor. As
such we compute $\hat{V}(Q^2)$ which is related to $V(Q^2)$
by\footnote{$\hat{V}(Q^2)$ is what would appear if we were computing
  the $\rho^+ \to \pi^+ \gamma$ transition in the isospin limit}
\begin{equation}
  \label{eq:vhat}
  V(Q^2) = 2 \times \frac{2}{3}e \times \hat{V}(Q^2),\nonumber
\end{equation}
and hence
\begin{equation}
  \label{eq:width_rhopi3}
  \Gamma(J/\psi \to \eta_c \gamma) =
  \alpha \frac{|\vec{q}|^3}{(m_{\eta_c} + m_\psi)^2}
  \frac{64}{27} \big|\hat{V}(0) \big|^2.
\end{equation}

There is only one direct experimental measurement of this width
\cite{Gaiser:1985ix}, $\Gamma_{\mathrm{CB}}(J/\psi \to \eta_c \gamma)
= 1.14(33) \mathrm{keV}$. There are, in addition, several measurements
of the product branching ratio for the process $J/\psi \to \eta_c \gamma
\to \phi \phi \gamma$\cite{PDBook} and one independent measurement of
$\eta_c \to \phi \phi$\cite{Huang:2003gc}. Taken together these data
give $\Gamma_{\phi\phi}(J/\psi \to \eta_c \gamma) = 2.9(1.5)
\mathrm{keV}$, which is consistent within the large errors with the
Crystal Ball result.

There is an essential ambiguity in how we compare our lattice results
with the experimental data which arises from the fact that our lattice
charmonia masses do not coincide exactly with the physical masses. The
problem lies in whether we should use experimental or lattice masses
in equation \eqref{eq:width_rhopi3}. $|\vec{q}|$ is closely related to
the hyperfine splitting which is rather sensitive to details of the
lattice calculation (as discussed in section \ref{spectrum}), hence we observe
considerable difference in using lattice or experimental masses:

\begin{tabular}{c|cc}
  $|\hat{V}(0)|_{\mathrm{expt.}}$ & Crystal Ball & ``$\phi \phi$'' \\
\hline
phys. masses & $1.27(19)$ & $2.02(52)$ \\
latt. masses & $1.56(22)$ & $2.48(64)$
\end{tabular}

\begin{figure}[h!]
  \centering
      \psfig{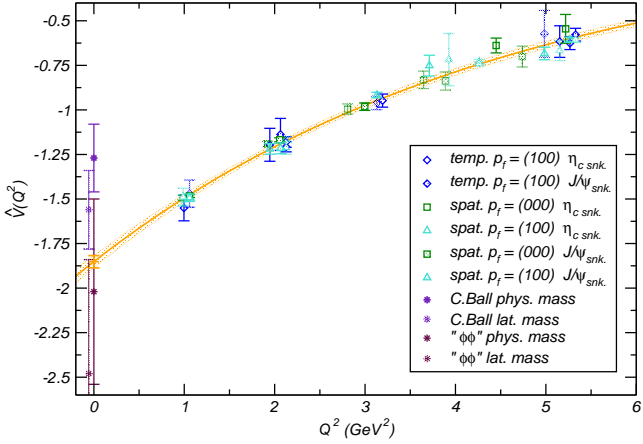}
  \caption{$J/\psi \to \eta_c \gamma$ transition form-factor}
  \label{fig:rho_pi}
\end{figure}

We extract $\hat{V}(Q^2)$ from three-point functions computing using
the following sequential sources at the sink: $\pi[\vec{p}_f =
(0,0,0); (1,0,0)]; \rho_{(x,y,z)}[\vec{p}_f =(0,0,0); (1,0,0)]$. The
values from different sink particle and momentum choices are all seen
to be consistent suggesting that the sequential source technique is
working correctly. The lattice data displayed in figure
\ref{fig:rho_pi} is fitted with
\begin{equation}
  \label{eq:rhopi_fit}
  \hat{V}(Q^2) = \hat{V}(0) e^{-\frac{Q^2}{16 \beta^2}}\nonumber
\end{equation}
 where we find
\begin{equation}
  \label{eq:rhopi_fit_param}
   \hat{V}(0) = 1.85(4); \quad \beta = 540(10) \mathrm{MeV}. \nonumber
\end{equation}
This agrees reasonably with the Crystal Ball result in the case of using lattice
masses throughout. Without performing a computation at the correct quark
mass and relaxing approximations sufficiently to duplicate the
experimental hyperfine splitting, we cannot give a more definitive
result.

The gaussian fit performed to the data was clearly rather successful;
if we wish we can motivate such a form by returning to the sort of
simple quark models considered in the previous section. Within these
models one performs a non-relativistic reduction of the current
operator and computes the matrix element of this between the $J/\psi$
and $\eta_c$ wavefunctions. This is a magnetic dipole ($M_1$)
transition that occurs through quark spin-flip - the lowest order
current operator (in an expansion in $v/c$) has the form $\vec{\sigma}
\times (\vec{p}\,'_q - \vec{p}_q)$. Computing in any frame (recalling
that the wavefunctions are assumed to be unchanged under boosts and
that we have the minimal non-relativistic reduction of the current) we
find
\begin{equation}
  V(Q^2) \propto \int r^2 dr R^*_{\eta_c}(r) j_0\big(\tfrac{|\vec{q}|
    r}{2}\big) R_{\psi}(r) \to \exp\Big[-\frac{|\vec{q}|^2}{16\bar{\beta}^2}\Big]\nonumber
\end{equation}
for harmonic oscillator wavefunctions. If the $\eta_c$ and $J/\psi$
are allowed different $\bar{\beta}$ then $\bar{\beta}^2 = \tfrac{1}{2}\big(
\bar{\beta}_{\eta_c}^2 + \bar{\beta}_\psi^2 \big)$. Once again we
encounter the problem that $V(Q^2)$ should be frame independent, but
$|\vec{q}|$ is not. In the rest frame of a decaying $J/\psi$, a
``natural'' frame to consider for this process,
\begin{equation}
 |\vec{q}|^2 = \frac{(m_\psi^2 - m_{\eta_c}^2)^2 +
  2 Q^2 ( m_\psi^2 + m_{\eta_c}^2) + Q^4  }{4 m_\psi^2}.\nonumber
\end{equation}
At small $Q^2$, $|\vec{q}|^2 \to \tfrac{(m_\psi^2 - m_{\eta_c}^2)^2}{4
  m_\psi^2} + Q^2 \tfrac{1}{2}\left( 1 +
\tfrac{m_{\eta_c}^2}{m_\psi^2}\right)$, which means the simple quark
model with harmonic oscillator wavefunctions has
the same $Q^2$-dependence as our fit. Other reasonable frames, in the
small $Q^2$ limit, have the same dependence.

Performing the full calculation (with $\kappa_c=0$) at $Q^2=0$ in a
quark model gives an expression
like\cite{Eichten:2002qv,Barnes:2005pb}
\begin{equation}
  \label{eq:M1}
  \Gamma = \alpha \frac{4 e_c^2}{3 m_c^2} |\vec{q}|^3 \left| \int r^2
    dr R^*_{\eta_c}(r) j_0\big(\tfrac{|\vec{q}|
    r}{2}\big) R_{\psi}(r)   \right|^2,\nonumber
\end{equation}
which can be used to evaluate the width provided the wavefunctions and
the mass of the charm quark are known. Determining this latter
quantity is a tricky issue - it is usually set at the Schr\"odinger
equation stage along with the phenomenological potential parameters
such that the mass spectrum is reproduced, however a unique solution
is not easily found and as such this mass parameter can vary
considerably between models, e.g. $m_c=1.84 \mathrm{GeV}$ in
\cite{Eichten:2002qv} while $m_c = 1.628 \mathrm{GeV}$ in
\cite{Godfrey:1985xj} and even $m_c=1.479 \mathrm{GeV}$ in
\cite{Barnes:2005pb}. Given this uncertainty, estimates for the $M_1$
transitions are rather imprecise, depending as they do, on the inverse
square of the mass\footnote{As mentioned earlier, with the pNRQCD EFT
approach one has a method to set the charm quark mass at a given order
in the power counting so this problem is
mollified ~\cite{Brambilla:2005zw}. A quark mass sensitivity can also
arise in the application of QCD sum-rules to this process; while this
can be lessened it has not proven possible to accommodate a value for
the transition amplitude as
low as experimentally found by Crystal Ball ~\cite{Beilin:1984pf}.   }. In principle we do not have this
problem in our lattice simulations - while we do vary the quark mass
to get agreement with the spectrum (we have to set the charm quark
mass somehow), we have no other free parameters to
vary. Unfortunately, as discussed in section \ref{spectrum} we did not
tune the quark mass perfectly and this is a source of some systematic
error on our result for this transition and may be the reason our
result is somewhat too large (recall that our quark mass is slightly
too small).

\subsection{$\chi_{c0} \to J/\psi \gamma_{E1, C1}$}
Experimentally it is only possible to access transverse on-shell
($Q^2=0$) photons in this transition and the matrix element is purely
through the electric dipole ($E_1$). In more generality, if we allow
$Q^2 \neq 0$ and necessarily also longitudinal photons, there is a
second multipole, labeled $C_1$. The decomposition of the transition
matrix element in terms of these multipoles is derived in the appendix
and we reproduce it here:
\begin{widetext}
\begin{align}
\langle S(\vec{p}_S) | j^\mu(0) | V(\vec{p}_V, r) \rangle =
\Omega^{-1}(Q^2) \Bigg( &E_1(Q^2) \Big[ \Omega(Q^2)
  \epsilon^\mu(\vec{p}_V, r) - \epsilon(\vec{p}_V, r).p_S \big(
  p_V^\mu p_V.p_S - m_V^2 p_S^\mu \big) \Big] \nonumber\\
  &\quad + \frac{C_1(Q^2)}{\sqrt{q^2}} m_V  \epsilon(\vec{p}_V, r).p_S \Big[ p_V.p_S (p_V+p_S)^\mu - m_S^2 p_V^\mu - m_V^2 p_S^\mu \Big] \Bigg).\nonumber
\end{align}
\end{widetext}
The Lorentz invariant matrix elements for the transition $\chi_{c0}
\to J/\psi \gamma^*(Q^2)$ are also given in the
appendix:
\begin{align}
{\cal M}(r_\gamma = \pm; r_\psi = \mp) &= E_1(Q^2) \nonumber\\
{\cal M}(r_\gamma = 0; r_\psi = 0) &= - C_1(Q^2).  \nonumber
\end{align}
Hence the analogue of \eqref{eq:width_rhopi} gives for the width at
$Q^2=0$,
\begin{equation}
  \label{eq:width_rhoa0}
  \Gamma(\chi_{c0} \to J/\psi \gamma) = \alpha
  \frac{|\vec{q}|}{m^2_{\chi_{c0}}} \frac{16}{9} \big|\hat{E_1}(0)\big|^2,\nonumber
\end{equation}
where the lattice form-factor is again related to the physical one by
$E_1(Q^2) = 2 \times \tfrac{2}{3} e \times \hat{E_1}(Q^2)$.

The most recent measurement of this decay's branching fraction comes
from the CLEO collaboration\cite{Adam:2005uh}, who find, using the PDG
total width to normalise: $\Gamma(\chi_{c0} \to J/\psi \gamma) =
204(31) \mathrm{keV}$. In addition to this we have the PDG\cite{PDBook}
average/fit to data obtained up to 2005 which gives $\Gamma(\chi_{c0}
\to J/\psi \gamma) = 115(14) \mathrm{keV}$. The next PDG report will
likely contain the CLEO value in a new average which will thus lie
between these two values.

\begin{figure}[h]
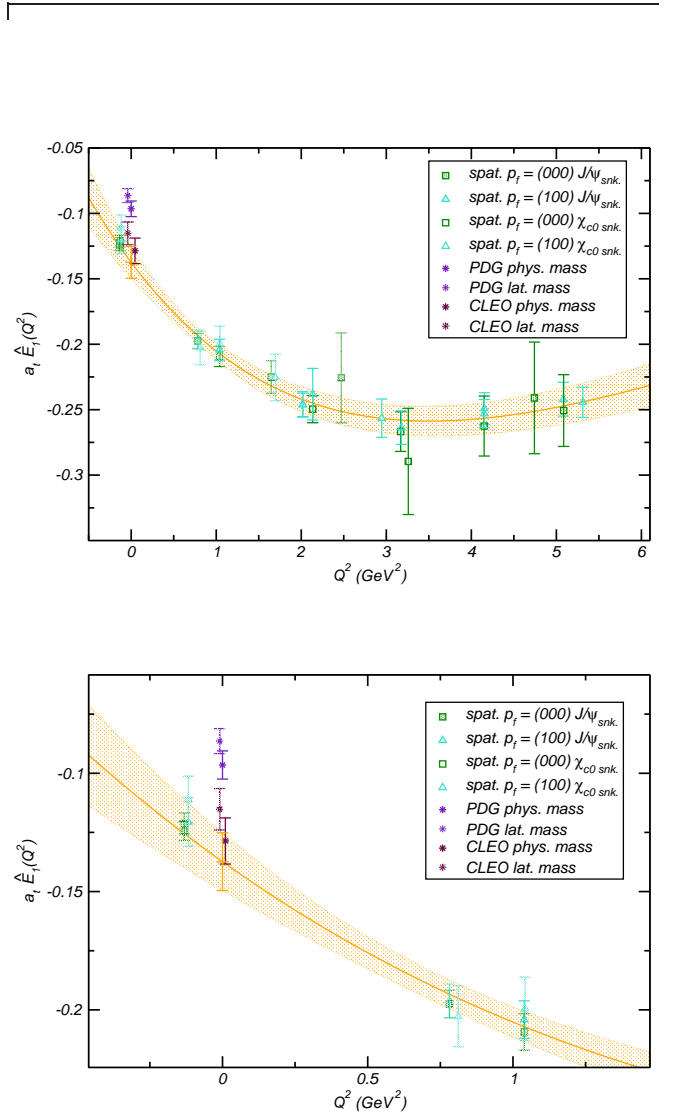

  \centering
      \psfig{width=8.5cm,file= rho-a0_E1.eps}
      \psfig{width=8.5cm,file= rho-a0_E1_zoom.eps}
  \caption{$\chi_{c0} \to J/\psi \gamma$ $E_1$ transition
    form-factor. (a) full range of lattice data (b) zoom to the $Q^2
    \approx 0$ region}
  \label{fig:rhoa0_E1_ff}
\end{figure}

In figure \ref{fig:rhoa0_E1_ff} we display the $\hat{E_1}(Q^2)$
extracted from our lattice simulations. Temporal vector current
insertions produce compatible results but with much larger error bars
and are not shown.

Our simulation data lies at $Q^2 \neq 0$, but since we are primarily
interested in the photopoint we require some fit function to allow us
to extrapolate back. In the light of the success of forms motivated by
the non-relativistic quark model in previous sections we consider
using a function which resembles one that would be derived in such a
model. We opt to use a form
\begin{equation}
  \label{eq:polyfit}
  \hat{E}_1(Q^2) = \hat{E}_1(0) \left( 1 + \frac{Q^2}{\rho^2} \right)
  \exp\left[ - \frac{Q^2}{16 \beta^2} \right],
\end{equation}
which has the gaussian behaviour used previously modified by a
polynomial in $Q^2$. In the simple quark model, the $Q^2/\rho^2$ term
could arise from relativistic corrections or departures from gaussian
wavefunction behaviour. Note that this form is analytic for $Q^2>0$ as
we would expect - singularities (as in the VMD case) will occur at
$Q^2 < 0$.
 
We do not include in the fit the points at $Q^2 < 0$ - these data,
corresponding to the case $\vec{p}_f = \vec{p}_i$ where $Q^2 = - (E_f
- E_i)^2$, were extracted from correlators with no plateau behaviour
using the fitting method described in section \ref{three}. It is
therefore a rather non-trivial cross-check that our fit function,
constrained by points at $Q^2 \gtrsim 1 \mathrm{GeV}^2$, extrapolated
to the $Q^2<0$ region, overlays these points.

The fit returns the following parameters:
\begin{gather}
  \label{eq:rhoa0_fit_param}
   a_t \hat{E}_1(0) = -0.137(12)\nonumber\\  
   \beta = 542(35)
   \mathrm{MeV}; \quad \rho = 1.08(13) \mathrm{GeV}\nonumber
\end{gather}

The longitudinal photon transition form-factor, $C_1(Q^2)$ can also be
extracted from lattice three-point functions. We display our data in
figure \ref{fig:rhoa0_C1_ff}.  As discussed in the appendix,
$\tfrac{C_1(Q^2)}{\sqrt{q^2}}$ is real for real $Q^2=-q^2$ if time-reversal
invariance holds - thus for $Q^2 > 0$, $C_1(Q^2)$ is imaginary. The
fit is to the function $i \tilde{c} \sqrt{Q^2} e^{-\tfrac{Q^2}{16
\beta^2}}$ which has the required property that $C_1(Q^2 \to
0) \to 0$ and which is what one would expect in a simple quark model.

\begin{figure}[h]
  \centering
      \psfig{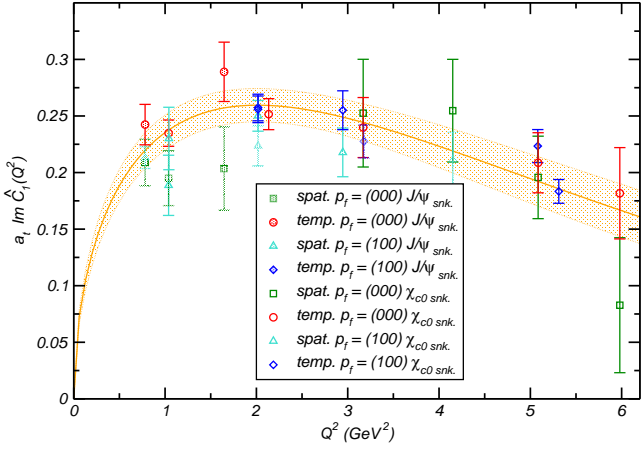}
  \caption{$\chi_{c0} \to J/\psi \gamma$ $C_1$ transition
    form-factor }
  \label{fig:rhoa0_C1_ff}
\end{figure}

The fit suffers from the large error bars on the data, but does at
least yield a $\beta$ value compatible with the value extracted from
the $E_1$ fit:
\begin{equation}
  \label{eq:rhoa0c1_fit_param}
   a_t \tilde{c}  = 1.83(16); \quad \beta = 501(33)
   \mathrm{MeV}\nonumber
\end{equation}

\subsection{$\chi_{c1} \to J/\psi \gamma_{E1, M2, C1}$}
With real photons this transition receives contributions from two
multipoles, the dominant electric dipole ($E_1$) and a much suppressed
magnetic quadrupole ($M_2$). Experimentally the $M_2$ contribution is
measured through the angular distribution of photons - the
PDG\cite{PDBook} average the two extractions
performed\cite{Ambrogiani:2001jw,Oreglia:1981fx}, each of which found
a number consistent with zero, to give
\begin{equation}
  \label{eq:m2}
  \frac{M_2(0)}{\sqrt{ E_1(0)^2+M_2(0)^2 }} = -0.002 \substack{+0.008\\-0.017}.\nonumber
\end{equation}

Appendix \ref{multipole} contains the tools required to derive the relation
connecting the transition matrix element with the multipole
amplitudes:
\begin{widetext}
\begin{align}
  &\langle A(\vec{p}_A, r_A) | j^\mu(0) | V(\vec{p}_V, r_V)\rangle = \tfrac{i}{4\sqrt{2} \Omega(Q^2)} \epsilon^{\mu\nu\rho\sigma} (p_A
  - p_V)_\sigma \times \nonumber \\
  & \quad\times \Bigg[ E_1(Q^2) (p_A+p_V)_\rho \Big( 2m_A
  [\epsilon^*(\vec{p_A}, r_A).p_V] \epsilon_\nu(\vec{p_V}, r_V)  + 2m_V [\epsilon(\vec{p}_V, r_V).p_A]
  \epsilon^*_\nu(\vec{p}_A, r_A)  \Big) \nonumber\\
& \quad\quad\quad+ M_2(Q^2) (p_A+p_V)_\rho \Big( 2m_A
  [\epsilon^*(\vec{p_A}, r_A).p_V] \epsilon_\nu(\vec{p_V}, r_V)  - 2m_V [\epsilon(\vec{p}_V, r_V).p_A]
  \epsilon^*_\nu(\vec{p}_A, r_A)  \Big) \nonumber\\
& \quad\quad\quad+ \frac{C1(Q^2)}{\sqrt{q^2}} \Big( - 4
\Omega(Q^2) \epsilon^*_\nu(\vec{p}_A, r_A) \epsilon_\rho( \vec{p}_V,
r_V)   \nonumber \\
&\quad\quad\quad\quad + (p_A+p_V)_\rho \Big[ (m_A^2-m_V^2 + q^2) [\epsilon^*(\vec{p}_A, r_A).p_V]\; \epsilon_\nu(\vec{p}_V, r_V) + (m_A^2 - m_V^2 -q^2) [\epsilon(\vec{p}_V, r_V).p_A] \; \epsilon^*_\nu(\vec{p_A}, r_A) \Big]         \Big)\Bigg].
\end{align}
\end{widetext}
In extracting these amplitudes from our lattice three-point functions
we are struck with the problem that our $\chi_{c1}$ signal (from the
operator $\gamma^i \gamma^5$) becomes noisy after relatively few
timeslices, in fact at roughly the same time that the two-point
function begins to plateau (see figure \ref{fig:meff}(b,c)). Because
of this it was only possible to extract a convincing signal for the
$\chi_{c1}$ with $\vec{p}_f=(000)$, which considerably limits the
number of three-point functions available to us for the transition
$\chi_{c1} \to J/\psi \gamma$. In addition the plateaux in the few
available correlators are borderline, so that for the lowest $Q^2$
point a fit of the form \eqref{eq:fit_correlator} was applied.

\begin{figure}[h]
  \centering
      \psfig{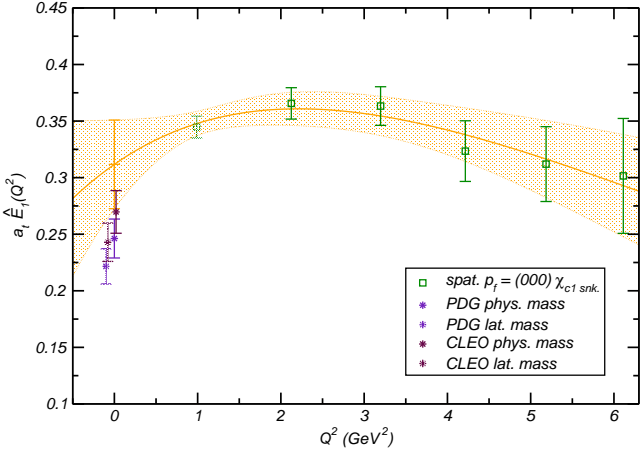}
  \caption{$\chi_{c1} \to J/\psi \gamma$ $E_1$ transition
    form-factor}
  \label{fig:rhoa1_E1_ff}
\end{figure}

In figure \ref{fig:rhoa1_E1_ff} we show the extracted $E_1(Q^2)$ and a
fit of the same form as used for $\chi_{c0} \to J/\psi \gamma_{E_1}$.
Within the large extrapolated error we are in agreement with the
experimental data. Unfortunately, unlike in the $\chi_{c0} \to J/\psi
\gamma$ case, the $E_1$ kinematical factor for $\vec{p}_f = \vec{p}_i$
where $Q^2 \approx 0$, is zero, so we cannot directly obtain $E_1(Q^2
\approx 0)$ and the subsequent cross-check on the extrapolation which
gave parameters,
\begin{gather}
  \label{eq:rhoa1e1_fit_param}
   a_t \hat{E}_1(0) = 0.312(39)\nonumber \\ 
\beta = 555(113)
   \mathrm{MeV}; \quad \rho = 1.65(59) \mathrm{GeV} \nonumber
\end{gather}

\begin{figure}[h]
  \centering
      \psfig{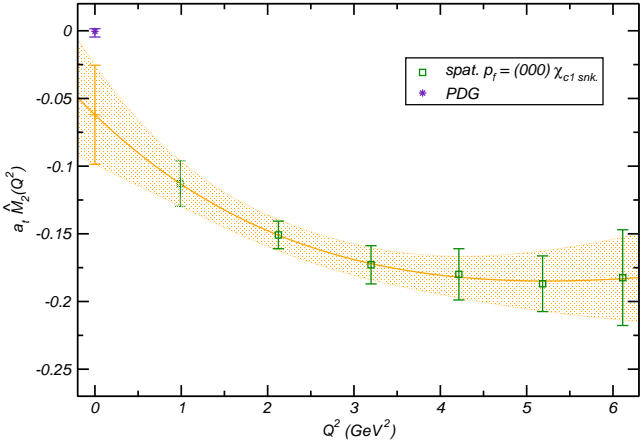}
  \caption{$\chi_{c1} \to J/\psi \gamma$ $M_2$ transition
    form-factor}
  \label{fig:rhoa1_M2_ff}
\end{figure}

In figure \ref{fig:rhoa1_M2_ff} we show the extracted $M_2(Q^2)$. The
fit has the same functional form as in the $E_1$ case and returns parameters:
\begin{gather}
  \label{eq:rhoa1m2_fit_param}
   a_t \hat{M}_2(0) = -0.062(37)\nonumber \\ 
\beta = 617(142)
   \mathrm{MeV}; \quad \rho = 0.93(47) \mathrm{GeV} \nonumber
\end{gather}
The $\beta$ value is compatible within the large error with the value extracted from the $E_1$ data.

We observe in figure \ref{fig:rhoa1_M2_ff} that our lattice data
extrapolates down to a value consistent with zero and hence with
experiment, within large errors, at $Q^2=0$.

There are a number of approaches we might consider to reduce the error
bar on the predicted value. The first is simply brute force; evaluate
on a larger number of gauge field configurations, thus reducing the
statistical fluctuations. Another option, which would also require
increased computation time, is to work with a larger spatial volume
(at the same lattice spacing); this allows smaller discrete
three-momenta and hence access to points closer to $Q^2=0$, the
reduced extrapolation distance then reducing the error on the
extrapolated point. A third possibility, ideally combined with the
previous two, is to enhance the $\chi_{c1}$ plateaux by finding an
interpolating field with maximal overlap on to the ground state
$\chi_{c1}$. This might involve diagonalising a matrix of two-point
functions in a basis of different smearings.

\begin{figure}[h]
  \centering
      \psfig{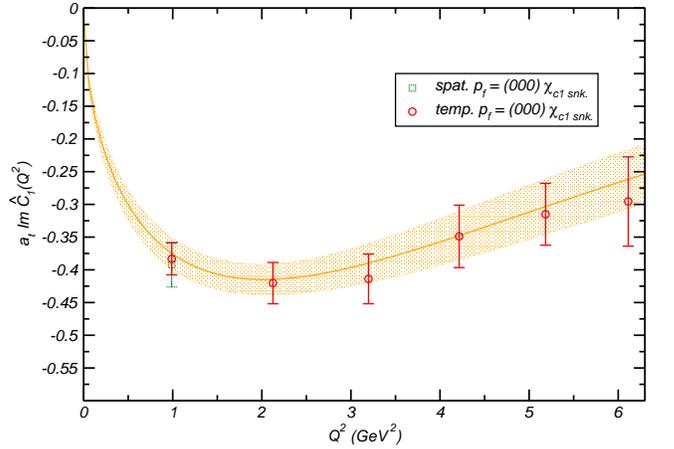}

  \caption{$\chi_{c1} \to J/\psi \gamma$ $C_1$ transition
    form-factor}
  \label{fig:rhoa1_C1_ff}
\end{figure}

Finally we have the longitudinal photon multipole $C_1$ which we
display in figure \ref{fig:rhoa1_C1_ff} along with a fit of same type
as used for the $\chi_{c0} \to J/\psi \gamma$ transition.
From the fit we extract
\begin{equation}
  \label{eq:rhoa1c1_fit_param}
   a_t \tilde{c}  = -2.91(26); \quad \beta = 502(38)
   \mathrm{MeV}\nonumber
\end{equation}

\subsection{$h_c \to \eta_c \gamma_{E1, C1}$}
The $h_c$, with $J^{PC}=1^{+-}$, was only recently observed with high
significance by the CLEO collaboration \cite{Rosner:2005ry, Rubin:2005px}. The reason
for the delay of discovery with respect to the other ground state
charmonia lies in the difficulty of production - the process
eventually utilised was the isospin violating $\psi(2S) \to \pi^0 h_c$
with a subsequent $h_c \to \eta_c \gamma$. Since only the product
branching fraction ${\cal B}(\psi(2S) \to \pi^0 h_c) {\cal B}(h_c \to
\eta_c \gamma)$ is measured, our calculation of $\Gamma(h_c \to \eta_c
\gamma)$ constitutes a prediction. 

\begin{figure}[h]
   \centering
       \psfig{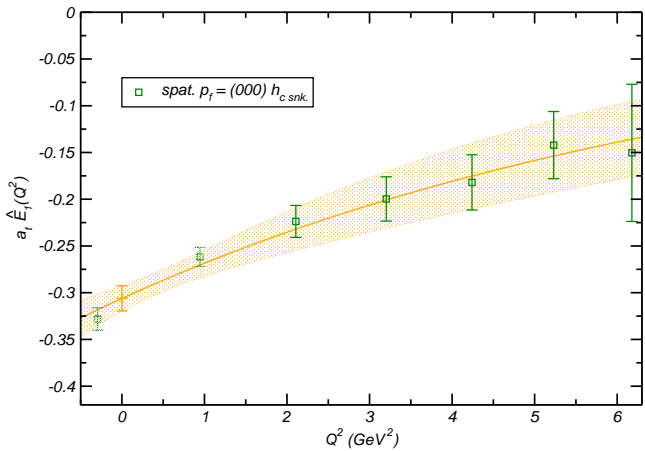}
   \caption{$h_c \to \eta_c \gamma$ $E_1$ transition form-factor}
   \label{fig:b1pi_E1_ff}
 \end{figure}

The form-factor decomposition is identical to the $\chi_{c0} \to
J/\psi \gamma$ case. Here again we suffer from poor two-point
functions - we were only able to extract a convincing $h_c$ signal in
the $\vec{p}_f=(000)$ case and as such we can use only three-point
functions with the $h_c$ at rest. Despite this limitation we are able
to extract some clean plateaux in a few cases and in the remaining
cases we fit using \eqref{eq:fit_correlator} - we plot the extracted
$E_1$ in figure \ref{fig:b1pi_E1_ff}. The fit shown (which is to all
points including the one at $Q^2 < 0$) is of the form
\begin{equation}
  \hat{E}_1(0) \exp - \frac{Q^2}{16 \beta^2},\nonumber
\end{equation}
since the data does not seem to require a $\tfrac{Q^2}{\rho^2}$ term
($\rho^{-1}$ in such a fit is very small). We find
\begin{equation}
  \label{eq:pib1e1_fit_param}
   a_t \hat{E}_1(0)  = -0.306(14); \quad \beta = 689(133)
   \mathrm{MeV}\nonumber
\end{equation}

The relation to the width is
\begin{equation}
  \label{eq:width_pib1}
  \Gamma(h_c \to \eta_c \gamma) = \alpha
  \frac{|\vec{q}|}{m^2_{h_c}} \frac{16}{27} \big|\hat{E_1}(0)\big|^2,\nonumber
\end{equation}
so that we predict
\begin{equation}
  \label{eq:width_pib1_number}
  \Gamma(h_c \to \eta_c \gamma) = \begin{matrix} 663(132)\\601(55) \end{matrix} \mathrm{keV}.\nonumber
\end{equation}
where the upper value uses lattice masses and the lower value physical masses.

\begin{figure}[h]
   \centering
       \psfig{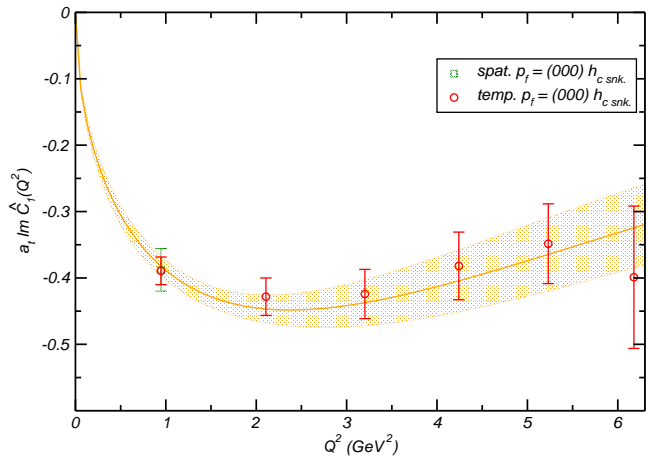}
   \caption{$h_c \to \eta_c \gamma$ $C_1$ transition form-factor}
   \label{fig:pib1_C1_ff}
 \end{figure}

The $C_1$ multipole is displayed in figure \ref{fig:pib1_C1_ff} with
the same fit form used for the $\chi_{c0} \to J/\psi \gamma$
transition.
\begin{equation}
  \label{eq:pib1c1_fit_param}
   a_t \tilde{c}  = -2.90(18); \quad \beta = 545(49)
   \mathrm{MeV}\nonumber
\end{equation}

\section{Discussion}\label{discuss}

We extracted, for the first time, a (limited) charmonium spectrum using
domain wall fermions on an anisotropic lattice. While an imperfect
quark mass tuning left our spectrum systematically $5\%$ too light,
the gross features of the $S$ \& $P$ levels were correct and notably
we found a rather large hyperfine splitting in contrast to other
quenched improved actions.

By attaching a vector current to only the quark line and not the
antiquark, we avoided the constraint of charge conjugation invariance
and sampled the vector form-factors of the lightest three charmonium
states. The charge form-factors of the $\eta_c$ and $J/\psi$ were
similar as expected, and, as one might anticipate on the basis of
dispersion relations, did not appear to be dominated by a single VMD
mechanism. 

For the $J/\psi$ the magnetic dipole form-factor was
extrapolated back to $Q^2=0$ to yield a magnetic dipole moment that was
consistent with there being zero anomalous charm quark magnetic moment
in a quark model picture. Similarly the very small quadrupole moment
extracted indicated minimal $D$-wave admixture into the $J/\psi$. The
$\chi_{c0}$ form-factor displayed a larger charge radius than the
$\eta_c$, indicating the effect of the centrifugal barrier in the
$P$-wave state.

In Table \ref{tab:transitions} we summarise our results for radiative
transitions, where it is clear that the $\beta$ values
for the $P \to S$ transitions are all compatible with having the same
value, which is in line with quark model expectations that the
$\chi_{cJ}, h_c$ spatial wavefunctions should differ only through
small spin-orbit distortions (and perhaps differing closed-channel
effects if these are allowed
for\cite{Eichten:1978tg,Eichten:1979ms}\footnote{But note that such
effects are not present within our quenched calculation}). Within
(p)NRQCD one might also have differing color octet
contributions\cite{Brambilla:2005zw}.

\begin{table}[h]
  \centering
  \begin{tabular}{c ccc}
    $\mathbf{E1}$ & $\chi_{c0} \to J/\psi \gamma$ & $\chi_{c1} \to
    J/\psi \gamma$ & $h_c \to \eta_c \gamma$ \\
    $\beta$/MeV & $542(35)$ & $555(113)$ & $689(133)$ \\
    $\rho$/MeV & $1080(130)$ & $1650(590)$ & $\infty$ \\
    $\Gamma\substack{\mathrm{lat. mass} \\ \mathrm{phys. mass}}$/keV &
    $\substack{288(60) \\ 232(41)}$ & $\substack{600(178) \\
      487(122)}$ & $\substack{663(132) \\ 601(55)}$ \\
$\Gamma\substack{\mathrm{PDG} \\ \mathrm{CLEO}}$/keV &
    $\substack{115(14) \\ 204(31)}$ & $\substack{303(44) \\
      364(31)}$ & -  \\
\hline
  \end{tabular}
  \begin{tabular}{c c||c c}
    $\mathbf{M1}$ & $J/\psi \to \eta_c \gamma$ &  $\mathbf{M2}$ & $\chi_{c1} \to
    J/\psi \gamma$ \\
     $\beta$/MeV & $540(10)$ &   $\beta$/MeV  & $617(142)$ \\
 $\Gamma\substack{\mathrm{lat. mass} \\ \mathrm{phys. mass}}$/keV &
    $\substack{1.61(7) \\ 2.57(11)}$ & $\tfrac{M2}{E1}$ &
    $-0.199(121)$ \\
$\Gamma\substack{\mathrm{PDG} \\ \mathrm{\phi\phi}}$/keV &
    $\substack{1.14(33) \\ 2.9(1.5)}$ & expt. &
    $-0.002(\substack{+8 \\ -17 })$ \\
\hline
  \end{tabular}
  \begin{tabular}{c ccc}
     $\mathbf{C1}$ & $\chi_{c0} \to J/\psi \gamma$ & $\chi_{c1} \to
    J/\psi \gamma$ & $h_c \to \eta_c \gamma$ \\
    $\beta$/MeV & $501(33)$ & $502(38)$ & $545(49)$ \\
    $|\tilde{c}|$/GeV & $11(1)$ & $17.6(1.6)$ & $17.5(1.1)$ \\
  \end{tabular}
  \caption{Radiative transitions}
  \label{tab:transitions}
\end{table}

We can also compare the pattern of $\rho$ values for the $E_1$
transitions with the expectations of a simple quark model. Performing
a non-relativistic reduction of the vector current, the $\rho$ term
arises from the spin-dependent correction ($\propto \vec{\sigma} \times
\vec{q}$) to the dominant convection current $\propto \vec{p}$; using
effective harmonic oscillator wavefunctions one finds\cite{Close:2002sf}
\begin{equation}
  E_1^{\mathrm{QM}}(Q^2) = a \left( 1 + r \frac{|\vec{q}|^2}{4
      \beta_\psi^2} \right) \exp - \frac{|\vec{q}|^2}{16
    \bar{\beta}^2},
\end{equation}
where $r$ is related to spin-orbit Clebsch-Gordan coefficients,
\begin{equation}
  r = \begin{matrix} 2 & \chi_{c0}\\ 1 & \chi_{c1}\\ 0 & h_c \end{matrix}.\nonumber
\end{equation}
Working in the $\chi_c$ rest frame at small $Q^2$ we would have
$|\vec{q}|^2 \approx |\vec{q}|_0^2 + Q^2 (1 + \Delta )$ where
$|\vec{q}|_0 = \tfrac{m_\chi^2 - m_\psi^2}{2 m_\chi}$ is the three-momentum
transfer at $Q^2=0$ and $\Delta = \tfrac{1}{2} \left(
  \tfrac{m_\psi^2}{m_\chi^2} -1\right)$. Thus we can express the quark
model form as
\begin{equation}
   E_1^{\mathrm{QM}}(Q^2) =  E_1^{\mathrm{QM}}(0) \left( 1 + r \frac{Q^2}{4
      \beta_\psi^2} \frac{1+\Delta}{1+\delta}  \right) \exp -
  \frac{Q^2 (1 + \Delta)}{16
    \bar{\beta}^2},
\end{equation}
with $\delta =  r \frac{|\vec{q}|_0^2}{4 \beta_\psi^2}$ and $
E_1^{\mathrm{QM}}(0) = a (1+ \delta) \exp - \frac{|\vec{q}|_0^2}{16
  \bar{\beta}^2} $. Hence, to a first approximation we'd expect that
$\rho \sim \tfrac{1}{\sqrt{r}}$ so that $\rho(\chi_{c1}) \approx
\sqrt{2} \rho(\chi_{c0})$ and $\rho(h_c) \to \infty$. In the same
approximation we have $\rho(\chi_{c0}) \approx 2 \beta_\psi$. Within
the large errors on the lattice results, these relations appear to be
satisfied.

Within the quark model, the $M_2$ transition is suppressed relative to
$E_1$ by one power of $v/c$. It is also rather sensitive to any charm
quark anomalous magnetic moment. Some details are worked out in
\cite{Sebastian:1992xq}, where they find a value (setting $\kappa_c =
0$) $M_2(0)/E_1(0) \sim -0.06$. Our data is unfortunately not sufficiently
accurate to discriminate on this level - we outlined earlier in the text some
possible improvements to the calculation to remedy this.

In this first attempt at charmonium radiative transitions using
lattice QCD we have demonstrated that it is possible to get reasonable
agreement with experiment and have gone some way to justifying certain
results of the more widely applied quark model. Future lattice work in this
direction will have to address the problem of reliable excited state
extraction in order to consider such well-measured transitions as
$\psi' \to \chi_{cJ} \gamma$. 

There is, naturally, a desire to see 
calculations done without the quenched approximation, but, as discussed in
section \ref{sim}, we do not expect unquenching to affect radiative
transitions particularly strongly, except in the sense that it will
improve the lattice state masses and help remove the phase-space
ambiguity we encountered in section \ref{trans}. However, an
unquenched computation is warranted to test models which propose a
considerable effect from coupled channels\cite{Eichten:1978tg}.

Our ultimate aim is to study photocouplings of light-quark hybrid mesons,
with this in mind the next step will be to consider radiative
transitions involving charmonium hybrids - the non-local interpolating
fields required for this study will also allow us to access higher
spin conventional charmonia such as the $\chi_{c2}$.

\begin{acknowledgments}
We thank Eric Swanson, Olga Lakhina, Frank Close, Ted Barnes and Jim
Napolitano for discussions.

This work was supported by DOE contract DE-AC05-84ER40150 under which
the Southeastern Universities Research Association operates the Thomas
Jefferson National Accelerator Facility.  Computations were performed
on clusters at Jefferson Laboratory under the SciDAC initiative.

\end{acknowledgments}

\appendix

\section{Multipole Decomposition}\label{multipole}
%derivation of Lorentz covariant multipole decompositions

It is convenient to express radiative transition amplitudes in terms
of multipoles. In this appendix we derive Lorentz covariant
decompositions of vector current matrix elements into
multipoles. These decompositions do not appear to have been explicitly
presented previously in the literature.

Our method involves writing down the most general Lorentz covariant,
current conserving and parity invariant decomposition of the matrix
element of the current in terms of a number of arbitrary
form-factors. We then compute the helicity amplitudes for the decay $i
\to f \gamma$ by contracting the current matrix element with a photon
polarisation vector. For convenience we work in a particular frame,
but the result is covariant. The relationship between helicity
amplitudes and multipoles is prescribed in \cite{Durand}, whence we
eliminate the arbitrary form-factors in favour of the multipole
form-factors.

We will demonstrate the method with the scalar-vector transition ($0^+
\leftrightarrow 1^-$) relevant to $\chi_{c0} \to J/\psi \gamma$. The
most general Lorentz covariant decomposition is:
\begin{widetext}
\begin{align}\label{SV}
  \langle S(\vec{p}_S)| j^\mu(0)| V(\vec{p}_V, r)\rangle& \nonumber \\
 &\hspace{-2cm}= A(q^2)
  \epsilon^{\mu\rho\sigma\tau} \epsilon_\rho(\vec{p}_V, r) p_{V\sigma}
  p_{S\tau} + P(q^2) \emu(\vec{p}_V, r) +   [\epsilon(\vec{p}_V,r).p_S] \Big[ B_+(q^2) (p_S+p_V)^\mu +  B_-(q^2) (p_S-p_V)^\mu \Big],
\end{align}
where the polarisation vectors carry a label $r$ which is the
$z$-component of the spin (which is not equal to the helicity in general).

Parity invariance requires that
\begin{eqnarray}\label{}
  \langle S(\vec{p}_S)| j^\mu(0)| V(\vec{p}_V, r)\rangle &=  \langle
  S(\vec{p}_S)| {\cal P}^{-1}{\cal P} j^\mu(0) {\cal P}^{-1}{\cal P} |
  V(\vec{p}_V, r)\rangle \nonumber \\
  &= - \mathsf{P}^\mu_\nu \langle S(-\vec{p}_S)| j^\nu(0)|
  V(-\vec{p}_V, r)\rangle, \nonumber 
\end{eqnarray}
where we've used ${\cal P}|S(\vec{p}_S)\rangle = |S(-\vec{p}_S)\rangle$,
${\cal P} |V(\vec{p}_V, r)\rangle = -|V(-\vec{p}_V, r)\rangle$ and
${\cal P}^{-1} j^\mu {\cal P} =
\mathsf{P}^\mu_\nu  j^\nu$  \footnote{$\mathsf{P}^\mu_\nu =
  g^{\mu\nu}= \mathrm{diag}(1,-1,-1,-1)$}. Hence
\begin{align}\label{}
    &A(q^2)
  \epsilon^{\mu\rho\sigma\tau} \epsilon_\rho(\vec{p}_V, r) p_{V\sigma}
  p_{S\tau} + P(q^2) \emu(\vec{p}_V, r) +  
   [\epsilon(\vec{p}_V,r).p_S] \Big[ B_+(q^2) (p_S+p_V)^\mu +  B_-(q^2)
   (p_S-p_V)^\mu \Big] \nonumber \\
&=  -\mathsf{P}^\mu_\nu \Big( A(q^2)
  \epsilon^{\nu\rho\sigma\tau} \epsilon_\rho(-\vec{p}_V, r) (\mathsf{P}p_V)_{\sigma}
  (\mathsf{P}p_S)_{\tau} +   P(q^2) \epsilon^\nu (-\vec{p}_V, r) 
  \nonumber \\   
 &\hspace{4cm} + [\epsilon(-\vec{p}_V,r).(\mathsf{P}p_S)] \Big[
 B_+(q^2) (\mathsf{P}p_S+\mathsf{P}p_V)^\nu +  B_-(q^2)
 (\mathsf{P}p_S-\mathsf{P}p_V)^\nu\Big] \Big). \nonumber
\end{align}
\end{widetext}
Properties of the rotation
group give that $\epsilon^\mu(-\vec{p}, r) = - \mathsf{P}^\mu_\nu
\epsilon^\nu(\vec{p}, r)$ using which one verifies the parity
invariance of this decomposition provided $A(q^2) = 0$.

The conservation of the vector current is an additional constraint,
\begin{align}
0 &= \partial_\mu  \langle S(\vec{p}_S)| j^\mu(x)| V(\vec{p}_V,
r)\rangle \nonumber \\
 &=  \partial_\mu \langle S(\vec{p}_S)| e^{i\hat{p}.x} j^\mu(0)
 e^{-i\hat{p}.x} | V(\vec{p}_V, r)\rangle \nonumber \\
&=  \partial_\mu e^{i(p_S-p_V).x} \langle S(\vec{p}_S)|  j^\mu(0) | V(\vec{p}_V,
r)\rangle\nonumber \\
\implies 0 &= q_\mu  \langle S(\vec{p}_S)| j^\mu(0)| V(\vec{p}_V,
r)\rangle,  \nonumber
\end{align}
whence we eliminate one of the three form-factors,
\begin{equation}\label{}
  P(q^2)=(m_V^2-m_S^2)B_+(q^2) - q^2 B_-(q^2)\nonumber.
\end{equation} 

Multipole amplitudes are most easily defined in terms of helicity
amplitudes which can be obtained from the decompositions of the
previous section in a straightforward way. First we will find the
defining relation for multipole amplitudes in terms of helicity
amplitudes. This is done in analogy with Durand \cite{Durand}, but rather than
working in the Breit frame we choose to work in the rest frame of the
decaying particle. The results are Lorentz covariant so this choice of
frame is irrelevant.

We can define a vertex function in the rest frame of a particle of spin-$J$ 
\begin{align}\label{vertex}
\Gamma^\nu(J'\lambda'; J\lambda) &\equiv \langle p' \hat{z};
J'\lambda'|j^\nu|J\lambda\rangle \nonumber\\
&= \langle J' \lambda' | e^{i \xi_{p'} K_3} j^\nu | J
\lambda\rangle,
\end{align}
where $\langle p' \hat{z}; J'\lambda'|$ is a final state of spin-$J'$,
\emph{helicity}-$\lambda'$ in motion along the positive $z$-axis with
momentum $p'$, $|J\lambda\rangle$ is a similar (initial) state at rest
and $e^{i \xi_{p'} K_3}$ is the unitary operator effecting the boost
from rest to momentum $p'\hat{z}$. The matrix element \eqref{SV}
discussed in the previous section is a generalisation of this vertex
functions to an arbitrary Lorentz frame,
i.e. $\Gamma^\mu(0,0;1,\lambda) = \langle
S(p'\hat{z})|j^\mu(0)|V(\vec{0}, r=\lambda)\rangle$ . Note that now we
are opting to use helicity and not the $z$-component of angular
momentum. However since we have chosen the end-state particle to move
along the positive $z$-direction, we have $\lambda = r, \lambda' =
r'$.

In this frame the amplitude for the decay $(J\lambda) \to
(J'\lambda')+(\gamma\lambda_\gamma)$  is
\begin{equation}\label{}
  {\cal M} = \epsilon^*_\nu( -p' \hat{z}, \lambda_\gamma) \Gamma^\nu(J'\lambda'; J\lambda),\nonumber
\end{equation}
where the polarisation vectors for a photon moving along the
$(-z)$-direction are
\begin{align}\label{}
  \epsilon^\mu(-p'\hat{z}, \lambda_\gamma = \pm) &= \pm
  \tfrac{1}{\sqrt{2}} (0,1,\mp i,0) \nonumber \\
 \epsilon^\mu(-p'\hat{z}, \lambda_\gamma = 0) &= \tfrac{1}{\sqrt{q^2}} (p',0,0, -\sqrt{p'^2+q^2}).\nonumber
\end{align}
We find that 
\begin{equation}\label{}
  {\cal M}(\lambda_\gamma) = \delta_{\lambda, \lambda'-\lambda_\gamma}    c_{\lambda_\gamma} \Gamma^{\lambda_\gamma}(J'\lambda';J\lambda),\nonumber
\end{equation}
where $c_\pm = 1, c_0 =-\tfrac{\sqrt{q^2}}{p'}$ and $\Gamma^{\pm}
\equiv \mp \tfrac{1}{\sqrt{2}}( \Gamma^1 \pm i \Gamma^2 )$.  

Returning to the definition of the vertex function \eqref{vertex}, we
assert that $e^{i \xi_{p'} K_3} j^\nu$ can be expressed as a sum of
operators which transform as tensors under the rotation group - this
is the essence of a multipole expansion: $e^{i \xi_{p'} K_3} j^\nu =
\sum_k T^k_{(\nu)}$. The vertex operator is then a sum of matrix
elements of these tensors, each of which satisfies the Wigner-Eckart
theorem, so that, for all spins integral, we can write
\begin{equation}\label{}
  \Gamma^\nu(J'\lambda'; J\lambda) = \sum_k (-1)^{J'+\lambda'}
  \begin{pmatrix} J' & k & J \\ -\lambda' & \nu & \lambda
  \end{pmatrix} \langle J' || T^k_{(\nu)} || J\rangle, \nonumber
\end{equation} 
where $\nu = \pm, 0$. $j^0$ and $j^\pm$ are in different
representations of the rotation group, so we'll allow different
reduced matrix elements for each. A purely conventional redefinition
of the reduced matrix element is $\langle J' || T^k_{(\pm)} ||
J\rangle = (-1)^{k} \sqrt{2k+1} \left[ E_k \tfrac{1}{2}(1+ (-1)^k
\delta P ) \mp M_k \tfrac{1}{2}(1- (-1)^k \delta P )\right]$ for the
transverse case and $\langle J' || T^k_{(0)} || J\rangle =
\tfrac{p'}{\sqrt{q^2}} (-1)^{k+1} \sqrt{(2k+1)} C_k \tfrac{1}{2}(1+
(-1)^k \delta P )$ for the longitudinal case \footnote{These
amplitudes are those in ~\cite{Sebastian:1992xq} and the experimental
study ~\cite{Ambrogiani:2001jw}}. $\delta P$ is the product of initial
and final meson parities and $E(M)$ indicates an electric(magnetic)
multipole. After some manipulation we obtain the multipole
decomposition of the helicity amplitudes,
\begin{widetext}
\begin{align}\label{}
  {\cal M}(\lambda_\gamma=\pm) &=  \sum_k \sqrt{\frac{2k+1}{2J+1}}
   \left[ E_k \tfrac{1}{2}(1+ (-1)^k \delta P ) \mp M_k
     \tfrac{1}{2}(1- (-1)^k \delta P )\right] \langle k\mp;
   J'\lambda\pm1 | J\lambda\rangle \nonumber \\
 {\cal M}(\lambda_\gamma=0) &= \sum_k \sqrt{\frac{2k+1}{2J+1}}  C_k \tfrac{1}{2}(1+ (-1)^k \delta P ) \langle k0; J'\lambda | J\lambda\rangle . \nonumber
\end{align}

For the particular case under study, $J=0, J'=1$; this transition has
one transverse multipole - $E_1$ and one longitudinal multipole -
$C_1$.
\begin{align}\label{SVmult}
 {\cal M}(\lambda_\gamma=\pm) &=    E_1(q^2) \nonumber\\
 {\cal M}(\lambda_\gamma=0) & =   - C_1(q^2) 
\end{align}
Contracting the complex conjugate of \eqref{SV} with a photon polarisation
vector yields, 
\begin{align}\label{SVarb}
 {\cal M}(\lambda_\gamma=\pm) &=  (m_V^2-m_S^2)  B_+(q^2) - q^2  B_-(q^2) \nonumber\\
 {\cal M}(\lambda_\gamma=0) &=  -\tfrac{\sqrt{q^2}}{2 m_V} \big[ (m_V^2-m_S^2+q^2) B_-(q^2) -  (3 m_V^2 + m_S^2 - q^2)  B_+(q^2) \big]
\end{align}
Solving \eqref{SVmult}, \eqref{SVarb} for $B_\pm(q^2)$ allows us to write the
current matrix element as a multipole expansion.

\begin{align}
\langle S(\vec{p}_S) | j^\mu(0) | V(\vec{p}_V, r) \rangle =
\Omega^{-1}(q^2) \Bigg( &E_1(q^2) \Big[ \Omega(q^2)
  \epsilon^\mu(\vec{p}_V, r) - \epsilon(\vec{p}_V, r).p_S \big(
  p_V^\mu p_V.p_S - m_V^2 p_S^\mu \big) \Big] \nonumber\\
  &\quad + \frac{C_1(q^2)}{\sqrt{q^2}} m_V  \epsilon(\vec{p}_V, r).p_S \Big[ p_V.p_S (p_V+p_S)^\mu - m_S^2 p_V^\mu - m_V^2 p_S^\mu \Big] \Bigg)
\end{align}

\end{widetext}

Note that this is expressed entirely in terms of invariants and
covariant quantities and hence can be used in any frame. One can check
that the tensor coefficients of the form-factors are orthogonal,
indicating the independence of $E_1, C_1$. The invariant quantity
$\Omega(q^2) \equiv (p_V.p_S)^2- m_V^2 m_S^2 = \tfrac{1}{4}
[(m_V-m_S)^2-q^2][(m_V+m_S)^2-q^2]$, and takes the simple value $m_S^2
|\vec{q}|^2$ in the rest frame of the decaying scalar.

The decomposition is further constrained by Minkowski time-reversal invariance
of the matrix element; this is a symmetry of the system if the meson
states are stable. Time-reversal is implemented as an anti-unitary
operation (${\cal T}$). As such the statement of
invariance for the matrix element of a hermitian operator is 
$\langle \beta | A | \alpha \rangle
= \langle \tilde{\alpha} | {\cal T} A {\cal T}^{-1} |
\tilde{\beta} \rangle = \big( \langle \tilde{\beta} | {\cal T}  A
{\cal T}^{-1} | \tilde{\alpha} \rangle)^* $ where $|\tilde{\alpha} \rangle
= {\cal T}|\alpha\rangle$. States of definite momentum, spin and
$z$-component of spin transform as ${\cal T} |\vec{p}, J, r \rangle =
\zeta (-1)^{J-r} |-\vec{p}, J, -r \rangle$, where $\zeta$ is an
arbitrary unphysical phase that we can choose independently for each
particle type\cite{Weinberg:1995mt}. For the scalar and vector we choose
$\zeta=+1$ - with this the vector has a real decay constant, $f_V$
defined in the standard way: $\langle 0 | \bar{\psi}(0) \gamma^\mu
\psi(0) | V(\vec{q}, r)\rangle = m_V f_V
\epsilon^\mu(\vec{q},r)$. 
Note that to get a real pseudoscalar decay constant
one needs to choose $\zeta_P = -1$ (see the next appendix).

The vector current transforms as
$ {\cal T} j^\mu  {\cal T}^{-1} = \mathsf{P}^\mu_\nu j^\nu$. Using the relation
$\epsilon^{\mu*}(\vec{p},r) = (-1)^{r+1} \mathsf{P}^\mu_\nu \epsilon^\nu(-\vec{p},
-r)$ one can show that $E_1(q^2)$ and $\tfrac{C_1(q^2)}{\sqrt{q^2}}$
are real.

An equivalent procedure for the axial-vector transition ($1^+
\leftrightarrow 1^-$) yields the decomposition
\begin{widetext}
\begin{align}
  &\langle A(\vec{p}_A, r_A) | j^\mu(0) | V(\vec{p}_V, r_V)\rangle = \tfrac{i}{4\sqrt{2} \Omega(q^2)} \epsilon^{\mu\nu\rho\sigma} (p_A
  - p_V)_\sigma \times \nonumber \\
  & \quad\quad \times \Bigg[ E_1(q^2) (p_A+p_V)_\rho \Big( 2m_A
  [\epsilon^*(\vec{p_A}, r_A).p_V] \epsilon_\nu(\vec{p_V}, r_V)  + 2m_V [\epsilon(\vec{p}_V, r_V).p_A]
  \epsilon^*_\nu(\vec{p}_A, r_A)  \Big) \nonumber\\
& \quad\quad\quad+ M_2(q^2) (p_A+p_V)_\rho \Big( 2m_A
  [\epsilon^*(\vec{p_A}, r_A).p_V] \epsilon_\nu(\vec{p_V}, r_V)  - 2m_V [\epsilon(\vec{p}_V, r_V).p_A]
  \epsilon^*_\nu(\vec{p}_A, r_A)  \Big) \nonumber\\
& \quad\quad\quad+ \frac{C1(q^2)}{\sqrt{q^2}} \Big( - 4
\Omega(q^2) \epsilon^*_\nu(\vec{p}_A, r_A) \epsilon_\rho( \vec{p}_V,
r_V)   \nonumber \\
&\quad\quad\quad\quad + (p_A+p_V)_\rho \Big[ (m_A^2-m_V^2 + q^2) [\epsilon^*(\vec{p}_A, r_A).p_V]\; \epsilon_\nu(\vec{p}_V, r_V) + (m_A^2 - m_V^2 -q^2) [\epsilon(\vec{p}_V, r_V).p_A] \; \epsilon^*_\nu(\vec{p_A}, r_A) \Big]         \Big)\Bigg]
\end{align}

\end{widetext}

\section{Minkowski and Euclidean N-point functions}\label{Mink}
We obtained all our Lorentz decompositions in Minkowski space, but
perform lattice computation in Euclidean; in this appendix we outline
a simple way to effect the mapping between
the two and describe some discrete symmetry constraints.

\subsection{Two-point functions}
In Minkowski space the following fermion bilinears are hermitian:
\begin{align}
  S(x) &= \bar{\psi}(x) \psi(x)\nonumber \\
  P(x) &=  \bar{\psi}(x)i \gamma^5 \psi(x)\nonumber \\
  V^\mu(x) &= \bar{\psi}(x) \gamma^\mu \psi(x) \nonumber\\
  A^\mu(x) &= \bar{\psi}(x) \gamma^\mu \gamma^5 \psi(x) \nonumber\\
  T^{\mu\nu}(x) &=  \bar{\psi}(x) \sigma^{\mu\nu} \psi(x). \label{hermit}
\end{align}
The transformation of spin-0 and spin-1 fields under time-reversal can
be written
\begin{align}\label{}
  {\cal T} \varphi(x) {\cal T}^{-1} &= \zeta_\varphi^*
  \varphi(-{\mathsf P}x); \nonumber \\
 {\cal T} v^\mu(x) {\cal T}^{-1} &= \zeta_v^* {\mathsf P}^\mu_\nu v^\nu(-{\mathsf P}x),\nonumber
\end{align}
with arbitrary phases $\zeta$. The choices in \eqref{hermit}
correspond to
\begin{equation}\label{}
  \zeta_S = +1;\quad \zeta_P=-1; \quad \zeta_V = +1;\quad \zeta_A = +1,\nonumber
\end{equation}
so that the decay constants defined in \eqref{eq:decay} are real:
\begin{align}\label{}
  i q^\mu f_P &= ({\mathsf P}^\mu_\nu \zeta_P \langle 0 | \bar{\psi}(0) \gamma^\nu \gamma^5 \psi(0)
  |P(-\vec{q})\rangle )^*\nonumber \\
 &= -i {\mathsf P}^\mu_\nu {\mathsf P}^\nu_\rho q^\rho f_P^*
 \zeta_P \nonumber\\
\implies f_P &= f_P^*, \nonumber \\
\vspace{3mm}
 m_V f_V \epsilon^\mu(\vec{q}, r) &= \big( {\mathsf P}^\mu_\nu \zeta_V
 (-1)^{1-r} \langle 0 |  \bar{\psi}(0) \gamma^\nu \psi(0) |
 V(-\vec{q}, -r)\rangle \big)^*\nonumber\\
&= \zeta_V (-1)^{1-r} m_V f_V^*  {\mathsf P}^\mu_\nu
\epsilon^{\nu*}(-\vec{q}, -r) \nonumber\\
\implies f_V &= f_V^*.\nonumber
\end{align}
In addition, the overlap factors $Z$ must be real:
\begin{align}\label{}
  Z_{P} (\vec{p}) &\equiv \langle 0 | \bar{u} i \gamma^5
  d(\vec{0}, 0) | P (\vec{p}) \rangle \nonumber\\
 &= \big( \langle 0 | {\cal T}^{-1} {\cal T} \bar{u}  i \gamma^5  d(\vec{0}, 0) {\cal T}^{-1}
  {\cal T}| P (\vec{p})\rangle \big)^* \nonumber\\
&= \big(   \langle 0 | (-1) \bar{u}  i \gamma^5 d(\vec{0}, 0) \zeta_P
| P (- \vec{p})\rangle \big)^* = Z_{P}^* (- \vec{p} ),\nonumber
\end{align}
since for the truly local operator, $Z$ is not actually a function of
$\vec{p}$ and where for a rotationally-invariant smeared operator
$Z=Z(|\vec{p}|)$.

The mapping between Minkowski and Euclidean space-times is effected by
the transformations,
\begin{align}\label{mink_euclid}
  t &\to - i \tilde{t} \nonumber\\
  \gamma^0 &\to \tilde{\gamma}_4 \nonumber\\
  \gamma^k &\to i \tilde{\gamma}_k \nonumber\\
  \gamma^5 &\to -\tilde{\gamma}_5,
\end{align}
where the tilded quantities are Euclidean (e.g. the gamma matrices
satisfy $\{\tilde{\gamma}_\mu, \tilde{\gamma}_\nu\} = 2\delta_{\mu\nu}$).

Consider as an example the vector two-point function in Minkowski
space,
\begin{align}
\Gamma^{(2) i j}(\vec{p}, t) &= \sum_{\vec{x}} e^{i \vec{p}.\vec{x}}
\langle \bar{\psi} \gamma^i \psi(\vec{x}, t)  \big[ \bar{\psi}
\gamma^j \psi(\vec{0}, 0) \big]^\dag \rangle \nonumber \\
  &= \sum_N \frac{(Z^{(N)}_{\psi})^2}{2 E^{(N)}_{\psi}}
  e^{-i E^{(N)}_{\psi} t} \sum_r \epsilon^i(\vec{p}, r)
  \epsilon^{*j}(\vec{p}, r)\nonumber \\
  &= \sum_N \frac{(Z^{(N)}_{\psi})^2}{2 E^{(N)}_{\psi}}
  e^{-i E^{(N)}_{\psi} t} \left( \delta_{ij} + \frac{p^i
      p^j}{(m^{(N)}_{\psi})^2}\right),\nonumber
\end{align}
where we inserted a complete set of states and used the Minkowski
space completeness expression for polarisation vectors $\sum_r
\epsilon^\mu(\vec{p}, r)  \epsilon^{*\nu}(\vec{p}, r) = \left( -g^{\mu\nu} + \frac{p^\mu
      p^\nu}{p^2}\right)$. In the lattice calculation we
  actually compute a Euclidean quantity:
\begin{align}
\tilde{\Gamma}^{(2)}_{ij}(\vec{p}, \tilde{t}) &= -  \Big\langle
\sum_{\vec{x}} e^{i \vec{p}.\vec{x}}\; \mathrm{ tr }\Big\{ \big(
\tilde{\gamma}_5 G(x,0)  \tilde{\gamma}_5 \big)^\dag \tilde{\gamma}_j
G(x,0) \tilde{\gamma}_i   \Big\} \Big\rangle \nonumber\\
&= \sum_{\vec{x}} e^{i \vec{p}.\vec{x}}
\langle \bar{\psi} \tilde{\gamma}_i \psi(\vec{x}, t)  \bar{\psi}
\tilde{\gamma}_j \psi(\vec{0}, 0) \rangle \nonumber \\
 &= (-i)^2 \Gamma^{(2) i j}(\vec{p}, -i \tilde{t}) \nonumber \\
&= - \sum_N \frac{(Z^{(N)}_{\psi})^2}{2 E^{(N)}_{\psi}}
  e^{- E^{(N)}_\psi \tilde{t}} \left( \delta_{ij} + \frac{p^i p^j}{(m^{(N)}_\psi)^2}\right),\nonumber
\end{align}
where with this derivation it is clear that all energies and momentum
should be interpreted as the usual real Minkowski variants. All that
we had to do was apply the mapping \eqref{mink_euclid}. Note that
we only ever consider polarisation vectors in Minkowski space, where
they are easily defined.

\subsection{Three-point functions}
In Minkowski space, inserting two complete sets of states gives
\begin{widetext}
\begin{align}\label{}
  \Gamma^{(3) f \Gamma i}(\vec{p}_f, \vec{q}; t_f, t) =   \sum_{f, i}
  & \frac{e^{-i E_f t_f} e^{-i(E_i - E_f) t } }{2 E_f(\vec{p}_f)\;2 E_i(\vec{p}_i) }  \nonumber  \\
&\langle 0 |
   \bar{\psi}\Gamma_f \psi(\vec{0}, 0) |f(\vec{p}_f, r_f)\rangle \langle f(\vec{p}_f, r_f) |
   \bar{\psi} \Gamma \psi(\vec{0}, 0)   |i(\vec{p}_i, r_i) \rangle\Big( \langle 0 |  \bar{\psi}\Gamma_i \psi(\vec{0},0) |  i(\vec{p}_i, r_i) \rangle \Big)^*. \nonumber
\end{align}
\end{widetext}
The Minkowski space-time Lorentz decomposition of the matrix
element was discussed in the previous appendix. We actually compute
the Euclidean variant,
\begin{equation}
 \tilde{\Gamma}^{(3)}_{f \Gamma i}(\vec{p}_f, \vec{q}; \tilde{t}_f, \tilde{t})
 = c_f c_\Gamma c_i \Gamma^{(3) f \Gamma i}(\vec{p}_f, \vec{q}; -i \tilde{t}_f, -i\tilde{t})\nonumber
\end{equation}
where e.g. $\tilde{\Gamma}_i = c_i \Gamma^i$. As an explicit example
consider the $\eta_c$ form-factor where $\Gamma_f=\Gamma_i= i \gamma^5$
and $\Gamma = \gamma^\mu$. Then using \eqref{mink_euclid} we have
\begin{align}
\Gamma^{(3) P \gamma^k P}(\vec{p}_f, \vec{q}; -i \tilde{t}_f,
-i\tilde{t}) &= \mathrm{Im}\, \tilde{\Gamma}^{(3)}_{P \gamma_k
  P}(\vec{p}_f, \vec{q}; \tilde{t}_f, \tilde{t}) \nonumber \\
\Gamma^{(3) P \gamma^0 P}(\vec{p}_f, \vec{q}; -i \tilde{t}_f,
-i\tilde{t}) &= - \mathrm{Re}\, \tilde{\Gamma}^{(3)}_{P \gamma_4
  P}(\vec{p}_f, \vec{q}; \tilde{t}_f, \tilde{t}). \nonumber
\end{align}
One should perform the mapping on each transition computed to ensure
that one extracts the correct complex component with the right sign.

\section{Scale setting on anisotropic lattices}\label{scale}

We outline how to set the scale of dimensionful quantities when one
has differing spatial and temporal lattice spacings. The lattice
action, written in terms of dimensionful quantities is $S=
\sum_{x^\mu} a_s^3 a_t \bar{\psi} Q \psi$. If we scale the fermion
fields by $\breve{\psi} = a_s^{3/2} \psi$ then we get a lattice spacing
independent action, $\sum_{x^\mu} \breve{\bar{\psi}} \breve{Q} \breve{\psi}$,
if $\breve{Q} = a_t Q$ where the anisotropy will appear in the spatial
derivative operator.

On the anisotropic lattice a bosonic field would have a mass term in
the action proportional to $\sum_{x^\mu} a_s^3 a_t m^2 \phi^2$, so
that if we scale the mass as $\breve{m} = a_t m$ (as we must since it
appears in Euclidean correlators like $e^{- m t}$), then we have to
scale the boson field as $\breve{\phi} = \sqrt{\tfrac{a_s^3}{a_t}}
\phi$.

If we define the creation/annihilation operators by the continuum
decomposition\footnote{if we wished to deal with a particle of integer
  spin $>0$ we would simply include the appropriate dimensionless
  polarisation tensor and the following logic would not be changed}
\begin{equation}
  \phi(x) = \int \frac{d^3\vec{p}}{(2\pi)^3} \frac{1}{2 E_{\vec{p}}}
    \left( \alpha^\dag(\vec{p}) e^{ip.x} + \alpha(\vec{p}) e^{-ip.x} \right),\nonumber
\end{equation}
and use $p_i = \frac{2 \pi}{L_i a_s} n_i$ (3-momenta must scale with $a_s$
as they appear in the dimensionless combination $\vec{p}.\vec{x}$) we
have
\begin{equation}
  \phi(x) = \sqrt{\tfrac{a_t}{a_s^3}}\breve{\phi}  = \sum_{\vec{n}_p}
  \frac{1}{L^3 a_s^3} \frac{a_t}{2 \breve{E}_{\vec{p}}}
   \left( \alpha^\dag(\vec{p}) e^{ip.x} + \alpha (\vec{p}) e^{-ip.x} \right),\nonumber
\end{equation}
so that the dimensionless creation/annihilation operator is $\breve{\alpha} =
\sqrt{\tfrac{a_t}{a_s^3}} \alpha$. Then since a single particle state
is defined by $|N(\vec{p})\rangle = \alpha^\dag | 0 \rangle$, we
define a dimensionless single particle state by
$|\breve{N}(\vec{p})\rangle = \breve{\alpha}^\dag | 0 \rangle = \sqrt{\tfrac{a_t}{a_s^3}}|N(\vec{p})\rangle $. It
is easy to check that the resolution of the identity takes the
following form
\begin{align}
  1 &= \sum_N \int \frac{d^3\vec{p}}{(2\pi)^3} \frac{1}{2 E_{\vec{p}}}
  |N(\vec{p})\rangle \langle N(\vec{p}) | \nonumber \\
&\to L^{-3}
  \sum_N \sum_{\vec{n}_p} \frac{1}{2 \breve{E}_{\vec{p}}} |\breve{N}(\vec{p})\rangle \langle \breve{N}(\vec{p}) |,\nonumber
\end{align}
which has no dependence on either lattice spacing. One can then show
by insertion of the complete set of states that the two-point function
calculated in a lattice simulation is, as expected,
\begin{equation}
 \breve{\Gamma}^{(2)} = \frac{\breve{Z}^2}{2 \breve{E}} e^{-\breve{t} \breve{E}}\nonumber
\end{equation}

The scaling of $Z$ is easily found
\begin{equation}
  Z e^{-i \vec{p}.\vec{x} } \equiv \langle 0 | \bar{\psi} \Gamma \psi(x)
  | N(\vec{p}) \rangle \to \frac{1}{\sqrt{a_s^3 a_t}} \langle 0 | \breve{\bar{\psi}} \Gamma \breve{\psi}(x)
  | \breve{N}(\vec{p}) \rangle,\nonumber
\end{equation}
so that the dimensionless $\breve{Z}$ that appears in the lattice simulation
is $\breve{Z} = \sqrt{a_s^3 a_t} Z$.

The pseudoscalar decay constant, $f_P$, has mass dimension $1$ and is defined
by
\begin{equation}
\langle 0 | \bar{\psi}(0) \gamma^\mu \gamma^5 \psi(0) |P(\vec{q})\rangle = i f_P q^\mu.\nonumber
\end{equation}
Taking the temporal component and scaling to dimensionless quantities
we have
\begin{equation}
\frac{1}{\sqrt{a_s^3 a_t}} \langle 0 | \breve{\bar{\psi}}(0) \gamma^0 \gamma^5 \breve{\psi}(0) |
  \breve{P}(\vec{q})\rangle = i f \breve{E}_{\vec{q}} a_t^{-1},\nonumber
\end{equation}
so that $\breve{f}_P = \sqrt{\frac{a_s^3}{a_t}} f_P$. Thus to obtain the
physical decay constant from the lattice value we calculate
\begin{equation}
  f = \xi^{-3/2} a_t^{-1} \breve{f}.\nonumber
\end{equation}
The same formula applies to the vector decay constant.

As an example of setting the scale of a dimensionful transition
form-factor factor, consider the scalar to vector transition:
\begin{equation}
  \langle S(\vec{p}_S) | \bar{\psi} \gamma^\mu \psi | V(\vec{p}_V, r)
  \rangle = E_1(Q^2) \epsilon^\mu(\vec{p}_V, r) + \hdots \nonumber
\end{equation}
The LHS scales to the dimensionless version leaving only one factor of
$a_t$ in the denominator and hence $\breve{E_1} = a_t E_1$.

\bibliography{/Users/dudek/physics/reference/master}

\end{document}